\documentclass[twocolumn,aps,prx,10pt]{revtex4-2}

\usepackage{amsmath}
\usepackage{amssymb}
\usepackage{wasysym}
\usepackage{graphicx}
\usepackage{hyperref}
\usepackage{newtxtext}
\usepackage[varvw]{newtxmath}
\usepackage{mathtools}
\usepackage[shortlabels]{enumitem}
\usepackage{placeins} 
\usepackage{bbold}
\usepackage{physics}

\hypersetup{
    colorlinks,
    linkcolor={blue},
    citecolor={blue},
    urlcolor={blue}
}

\graphicspath{{./}{./figs/}}

\newcommand{\rme}{\mathrm{e}}
\newcommand{\rmi}{\mathrm{i}}
\newcommand{\rmd}{\mathrm{d}}

\makeatletter
\let\MyIntOrig\int
\def\MyIntSpace{\hspace{-.25em}} 
\def\int{\MyInt}
\def\MyInt{\MyIntOrig\MyIntSkipMaybe}
\def\MyIntSkipMaybe{
  \@ifnextchar_{\MyIntSkipScript}{%
  \@ifnextchar^{\MyIntSkipScript}{%
  \@ifnextchar\limits{\MyIntSkipTok}{%
  \@ifnextchar\nolimits{\MyIntSkipTok}{%
  \MyIntSpace}}}}%
}
\def\MyIntSkipScript#1#2{#1{#2}\MyIntSkipMaybe}
\def\MyIntSkipTok#1{#1\MyIntSkipMaybe}
\makeatother

\newcommand{\appsection}[1]{\section{\MakeUppercase{#1}}}

\begin{document}

\title{Dirac quantum criticality in twisted double bilayer transition metal dichalcogenides}

\author{Jan Biedermann}
\author{Lukas Janssen}

\affiliation{Institut f\"ur Theoretische Physik and W\"urzburg-Dresden Cluster of Excellence ct.qmat, TU Dresden, 01062 Dresden, Germany}

\begin{abstract}
We investigate the phase diagram of moiré double bilayer transition metal dichalcogenides with ABBA stacking as a function of twist angle and applied pressure. At hole filling $\nu = 2$ per moiré unit cell, the noninteracting system hosts a Dirac semimetal with graphene-like low-energy bands in the moiré Brillouin zone. At small twist angles, the Fermi velocity is reduced and interactions dominate the low-temperature behavior.
A strong-coupling analysis identifies insulating ferromagnetic and antiferromagnetic ground-state candidates, characterized by spin-density modulations set by the moiré scale. Using a realistic continuum model with long-range Coulomb interactions, we perform self-consistent Hartree-Fock calculations to study the competition between these states.
Varying the twist angle or pressure drives a transition from a Dirac semimetal to an antiferromagnetic insulator, which breaks SU(2) spin rotation and two-fold lattice rotation symmetries. Within a renormalization group analysis of the most general symmetry-allowed low-energy field theory, we show that this semimetal-to-insulator transition is continuous and belongs to the (2+1)D relativistic Gross-Neveu-Heisenberg universality class with $N = 2$ four-component Dirac fermions. Finite heterostrain, relevant in realistic samples, induces a crossover from Gross-Neveu-Heisenberg universality at intermediate temperatures to conventional (2+1)D Heisenberg criticality at the lowest temperatures.
Further decreasing the twist angle can cause a level crossing from the antiferromagnetic insulator into a ferromagnetic insulator with spin-split bands.
Our results provide a comprehensive theoretical framework that complements and elucidates recent experiments in twisted double bilayer WSe$_2$.
\end{abstract}

\date{January 30, 2026}

\maketitle

\section{Introduction}

A quantum phase transition occurs at absolute zero temperature, driven by a nonthermal control parameter such as magnetic field, pressure, or strain. 
When continuous, it gives rise to an extended quantum critical regime dominated by collective phenomena at finite temperatures above the transition point~\cite{sachdevbook}.
In many systems, this regime is governed solely by fluctuations of an order-parameter field and the physics can often be mapped onto those of a higher-dimensional classical phase transition.
More intriguing many-body phenomena, however, may arise when additional low-energy degrees of freedom are present, in which case the simple quantum-to-classical mapping no longer applies.

Quantum critical points involving Dirac fermions arise across multiple domains, bridging phenomena between condensed-matter and high-energy physics.
They have been studied in the context of graphene~\cite{herbut06,herbut09,assaad13,janssen14,otsuka16,pujari16,ray18,ray21b}, topological insulators~\cite{lee07,grover14}, frustrated quantum magnets~\cite{seifert20,ray21a,liu22,liu24,ray24,fornoville25}, and $d$-wave superconductors~\cite{vojta00,huh08,schwab22}, as well as in toy models of chiral symmetry breaking and nonperturbative renormalizability~\cite{hands93,gies10,braun11,janssen12,gehring15,dabelow19,hands19,cresswell23}.
In such systems, gapless Dirac fermions can acquire a dynamical mass gap through interactions, giving rise to nontrivial quantum critical behavior that has been extensively explored theoretically~\cite{boyack21,herbut24}.
For a long time, this physics remained purely theoretical and beyond experimental reach. Very recently, however, it has become possible to realize and probe Dirac quantum critical points in moiré materials tuned by twist angle. Theoretical studies of twisted bilayer graphene predicted a continuous transition between a Dirac semimetal and a Kramers intervalley-coherent insulator~\cite{biedermann25,huang25}, belonging to the Gross-Neveu-XY universality class~\cite{parthenios23, hawashin25}. At the same time, experiments on twisted double bilayer WSe$_2$~\cite{ma24} and twisted bilayer MoSe$_2$~\cite{yang25} have observed a Dirac-semimetal-to-insulator transition, interpreted as a realization of a related type of Dirac quantum critical point.

\begin{figure}[b!]
\includegraphics[width=\linewidth]{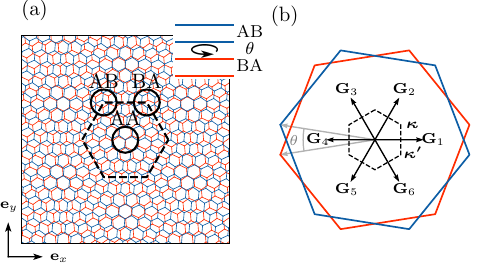}
\caption{
(a)~Real-space view of the two inner layers of ABBA-stacked twisted double bilayer TMDs, highlighting the moiré unit cell (dashed hexagon). Locally, regions of approximate AA, AB, and BA stacking can be identified. The inset illustrates the out-of-plane structure: an AB-stacked bilayer atop a BA-stacked bilayer, with a small twist angle $\theta$ between them.
(b)~The Brillouin zones of the top (blue hexagon) and bottom (red hexagon) bilayers are rotated relative to each other by a twist angle $\theta$, producing a moiré Brillouin zone (small dashed hexagon) with high-symmetry points $\boldsymbol{\kappa}$ and $\boldsymbol{\kappa}'$ at its corners. Neighboring moiré Brillouin zones are connected by the moiré reciprocal lattice vectors $\mathbf{G}_j$.
}
\label{fig:1}
\end{figure}

In this work, we theoretically investigate the phase diagram of hole-doped double bilayer transition metal dichalcogenides (TMDs) with a small twist angle between the two bilayers, see Fig.~\ref{fig:1}(a).
TMDs are layered materials with chemical formula $MX_2$, where $M$ is a transition metal, taken here to be a group-VI element such as Mo or W, and $X$ is a chalcogen such as S, Se, or Te. 
Earlier studies have placed particular emphasis on bilayer TMDs~\cite{devakul21, li21, zhang21, klebl23, saigal24, yang24, xu24, guerci25}. In such systems, however, the location of the valence-band maximum depends sensitively on the specific choice of metal and chalcogen~\cite{angeli21, pei22, gatti23}, making it challenging to realize a Dirac quantum critical point with SU(2) spin symmetry due to the spin splitting induced by spin-orbit coupling.
In contrast, the stronger interlayer hybridization in trilayer and double bilayer TMDs is generally expected to pin the valence-band maximum at the Brillouin-zone center $\boldsymbol{\Gamma}$~\cite{movva18, foutty23}, analogous to the situation in bulk TMDs~\cite{manzeli17}. The bands near $\boldsymbol{\Gamma}$ are spin degenerate, giving rise to an SU(2)-symmetric effective low-energy description.
Here we focus on ABBA-stacked double bilayer TMDs. Introducing a small twist between the two bilayers generates a tunable emergent honeycomb lattice that retains the SU(2) spin symmetry and two-fold lattice rotation symmetry of monolayer graphene, now realized on the moiré scale~\cite{pan23}.
Note that the crystal structure of ABBA-stacked double bilayer TMDs~\cite{pan23,ma24} differs from that of ABAB-stacked double bilayers~\cite{foutty23}, even when a finite twist is applied between the two inner layers, due to the absence of the two-fold rotation symmetry $C_{2y}$ about the in-plane $y$ axis indicated in Fig.~\ref{fig:1}(a).
In the noninteracting limit, the electronic spectrum hosts two gapless Dirac cones located at the moiré Brillouin-zone corners $\boldsymbol{\kappa}$ and $\boldsymbol{\kappa}’$, see Fig.~\ref{fig:1}(b).
At a hole filling of $\nu = 2$ per moiré unit cell, the Fermi level sits precisely at the band crossing, with no additional low-energy electronic states away from the $\boldsymbol{\kappa}$ and $\boldsymbol{\kappa}’$ points.
To investigate the low-temperature behavior of the system, we use a realistic continuum model with long-range Coulomb interactions, employing parameters relevant to experiments on double bilayer WSe$_2$~\cite{ma24}.
Our model consistently incorporates the angle dependence of the interaction strength and includes contributions from remote bands. It also enables the study of uniaxial pressure and residual heterostrain, as expected in realistic samples.
At large twist angles, the Dirac semimetal remains stable for realistic interactions. Decreasing the twist angle, or applying uniaxial pressure at sufficiently small angles, destabilizes the semimetal and induces a transition to an antiferromagnetic insulator with broken SU(2) spin and inversion symmetries, featuring a spin-density wavelength on the scale of the moiré lattice constant.
The semimetal-to-insulator transition is continuous and falls within the (2+1)D relativistic Gross-Neveu-Heisenberg universality class with $N = 2$ four-component Dirac fermions.
Finite heterostrain drives a crossover from Gross-Neveu-Heisenberg universality at intermediate temperatures to conventional (2+1)D Heisenberg criticality at ultralow temperatures.
At very small twist angles, the antiferromagnet competes with an insulating state exhibiting a finite total magnetization while preserving inversion symmetry. For sufficiently strong Coulomb interactions, decreasing the twist angle can drive a second transition from the antiferromagnetic insulator to this ferromagnetic insulator. This antiferromagnet-to-ferromagnet transition is found to be discontinuous.
While our calculations are performed using a model relevant for double bilayer WSe$_2$, we expect the qualitative features of our results to apply more broadly to twisted TMDs that host an emergent honeycomb lattice on the moiré scale.

The remainder of the paper is structured as follows. We begin in Sec.~\ref{sec:model} with a description of the model and outline our Hartree-Fock approach in Sec.~\ref{sec:hartree-fock}. The strong-coupling analysis presented in Sec.~\ref{sec:strong-coupling} sets the stage for the twist-angle-tuned phase diagram discussed in Sec.~\ref{sec:phase-diagram}. We then explore the influence of uniaxial pressure and heterostrain in Secs.~\ref{sec:pressure} and \ref{sec:strain}, respectively, before turning in Sec.~\ref{sec:critical} to the universal properties of the Dirac quantum critical point. We conclude in Sec.~\ref{sec:conclusions}. Technical details are deferred to three appendices.

\section{Interacting continuum model}
\label{sec:model}

\subsection{Kinetic terms}

To model the kinetic part of the electronic Hamiltonian in the presence of the moiré potential, we employ a continuum description of hole-doped ABBA-stacked twisted double bilayer TMDs with valence band edge at the $\boldsymbol \Gamma$ point in the Brillouin zone~\cite{angeli21, pan23}
\begin{align} \label{eq:continuum-model}
\mathcal{H}_{\mathrm{kin}} &= \sum_{\mathbf q} f^\dagger_{\mathbf q} \left(h_{\boldsymbol{\Gamma}}(\mathbf q) + h_0 \right) f_{\mathbf q} +
\sum_{\mathbf q} \sum_{j = 1}^6 f^\dagger_{\mathbf q + \mathbf G_j} h_1(\mathbf G_j) f_{\mathbf q},
\end{align}
with the $\boldsymbol \Gamma$-valley hole-band dispersion given by
\begin{align} \label{eq:gamma-dispersion}
h_{\boldsymbol{\Gamma}}(\mathbf q) &= - \frac{\hbar^2 \mathbf q^2}{2 m_{\boldsymbol \Gamma}} \mathbb{1}_4.
\end{align}
Throughout this work, we adopt the convention that $\mathbf q = \mathbf k + \mathbf G$ denotes unrestricted momenta, where $\mathbf k$ lies within the moiré Brillouin zone and $\mathbf G$ is a moiré reciprocal lattice vector.
The fermionic operators $f_{\mathbf q} = (f_{\mathbf q, \ell, s})$ annihilate electrons with wavector $\mathbf q$, layer index $\ell = 1,\dots,4$, and spin index $s = {\uparrow, \downarrow}$.
Following Refs.~\cite{movva18, foutty23, pan23}, we take the $\boldsymbol{\Gamma}$-valley hole effective mass in units of the free electron mass $m_\text{e}$ to be $m_{\boldsymbol \Gamma} = 1.2 m_\text{e}$.
$h_0$ and $h_1(\mathbf G_j)$ are $4 \times 4$ matrices acting in layer space. We assume the moiré potential to only affect electrons in the two inner layers, which are rotated by a small twist angle $\theta$ relative to each other.
The matrix $h_0$ is parametrized by four material-dependent constants $(V_1, V_{2}^{(0)}, V_{12}, V_{23}^{(0)})$. These describe the intralayer potentials in the two outer layers ($V_1$), the homogeneous component of the intralayer potentials in the two inner layers ($V_{2}^{(0)}$), the interlayer tunneling between outer and inner layers ($V_{12}$), and the homogeneous component of the tunneling between the two inner layers ($V_{23}^{(0)}$).
The matrix $h_1(\mathbf{G}_j)$ depends on the reciprocal lattice vectors $\mathbf{G}_j$, $j = 1, \dots, 6$, which connect neighboring moiré Brillouin zones [see Fig.~\ref{fig:1}(b)]. It involves three additional constants $V_{2}^{(1)}$, $\varphi$, and $V_{23}^{(1)}$, which describe the oscillating component of the intralayer potentials in the two inner layers ($V_{2}^{(1)}$), the corresponding phase shift ($\varphi$), and the oscillating component of the tunneling between the two inner layers ($V_{23}^{(1)}$).
The specific parameters adopted for modeling WSe$_2$ are provided in Appendix~\ref{app:model}.
Since spin-orbit coupling is negligible near the $\boldsymbol{\Gamma}$ point of the Brillouin zone, $h_{\boldsymbol{\Gamma}}$, $h_0$, and $h_1$ are all spin-diagonal.

The continuum Hamiltonian above captures the electronic excitations near the $\boldsymbol{\Gamma}$-valley valence band edge, providing a valid low-energy description at small hole doping.
Consistent with bilayer TMDs with $\boldsymbol{\Gamma}$-valley valence band edge~\cite{angeli21}, we find that the bandwidth of the two topmost moiré bands increases monotonically with twist angle. This behavior contrasts with twisted bilayer graphene, where the bandwidth instead exhibits a minimum at the magic angles~\cite{bistritzer11, tarnopolsky19}.
We focus on a filling of $\nu = 2$ holes per moiré unit cell and consider twist angles $\theta$ between $1^\circ$ and $4^\circ$, where the bandwidth remains small enough for correlated physics to emerge and the continuum model is expected to remain valid.

\subsection{Coulomb interaction}

The Coulomb interaction between electronic excitations is most conveniently expressed in the band eigenbasis of $\mathcal{H}_{\mathrm{kin}}$, using the fermionic operators $c_{\mathbf{k}} = (c_{\mathbf{k}, n, s})$, where $n$ labels the moiré band index~\cite{bultinck20, liu21, hofmann22, faulstich23, biedermann25}.
The interaction part of the Hamiltonian can then be written in momentum space as
\begin{align}
\mathcal{H}_{\mathrm{int}} &= - \frac{1}{2A}\sum_{\mathbf q} V_{\mathbf q} :\mathrel{\rho_{\mathbf q} \rho_{-\mathbf q}}:,
\end{align}
where $A = \sqrt{3} a_0^2 L^2 / 8 \sin^2(\theta / 2)$ is the area of a sample containing $L \times L$ moiré unit cells, and $a_0$ is the lattice constant of the monolayer TMD. For our calculations, we take $a_0 = 0.328\,\mathrm{nm}$, appropriate for WSe$_2$~\cite{schutte87}.
The operator $\rho_{\mathbf q} = \sum_{\mathbf k} c^\dagger_{\mathbf k} \Lambda(\mathbf k, \mathbf q) c_{\mathbf k + \mathbf q}$ represents the Fourier-transformed charge density, where $\Lambda_{m, s; n, s^\prime}(\mathbf k, \mathbf q) = \braket{u_{\mathbf k, m}}{u_{\mathbf k + \mathbf q, n}} \delta_{s, s^\prime}$ are the overlap matrices of the noninteracting eigenstates $\ket{u_{\mathbf k, n}}$.
We consider a screened repulsive Coulomb interaction in the presence of top and bottom metallic gates, in the limit of negligible sample thickness, given by
\begin{align} \label{eq:coulomb}
V_{\mathbf q} = \frac{e^2}{2 \epsilon_0 \epsilon_\text{eff} \left\vert \mathbf q \right\vert} \tanh(\vert \mathbf q \vert d),
\end{align}
where $e$ is the free electron charge and $\epsilon_\text{eff}$ is an effective relative dielectric permittivity accounting for both gate screening and internal screening arising from processes within the moiré bands. Such internal screening effects is not captured by the standard Hartree-Fock approach and must therefore be included manually~\cite{goodwin19, goodwin20a, goodwin20b, biedermann25, sanchez25}. While screening generally produces a momentum-dependent permittivity, previous studies on twisted bilayer graphene have shown that using the $\mathbf q \to 0$ limit of $\epsilon_\text{eff}(\mathbf q)$ yields qualitatively consistent results~\cite{biedermann25}.
For flat bands, internal screening can be substantial, with $\epsilon_\text{eff}(\mathbf q \to 0)$ reaching values up to $\sim 200$~\cite{goodwin19, biedermann25}.
To simplify the discussion, we treat $\epsilon_\text{eff}$ as a free parameter, neglecting its momentum and frequency dependence, and compute the twist-angle-dependent phase diagram for various values of $\epsilon_\text{eff}$. For comparison with experiments on samples with small twist angles between $1^\circ$ and $4^\circ$, values of $\epsilon_\text{eff} \sim \mathcal O(100)$ are considered realistic~\cite{goodwin19, biedermann25, munoz25}.
For the explicit calculations, the distance between the sample and each metallic gate is set to $d = 12.5\,\mathrm{nm}$, roughly corresponding to the devices studied in Ref.~\cite{ma24}.

\subsection{Full Hamiltonian and symmetries}

In the band basis, the full interacting continuum model can be written as
\begin{align} \label{eq:full-model}
\mathcal H = \tilde{\mathcal H}_{\mathrm{kin}} + \mathcal{H}_{\mathrm{int}},
\end{align}
where $\tilde{\mathcal{H}}_{\mathrm{kin}} = \sum_{\mathbf k} c^\dagger_{\mathbf k} h(\mathbf{k}) c_{\mathbf{k}}$ is the effective kinetic Hamiltonian in the band basis.
The matrix $h(\mathbf{k})$ incorporates a subtraction term to avoid double counting of interaction effects~\cite{bultinck20, faulstich23, kwan22}, as described in detail in Appendix~\ref{app:model}.

Because spin-orbit coupling is negligible in the $\boldsymbol{\Gamma}$ valley, the effective low-energy Hamiltonian preserves full $\mathrm{SU}(2)$ spin-rotation symmetry. 
The model further possesses time-reversal symmetry $\mathcal{T}$, a threefold rotation symmetry $C_{3z}$ around the out-of-plane $z$ axis, and a twofold rotation symmetry $C_{2y}$ around the in-plane $y$ axis.
The latter relates the AB and BA stacking regions indicated in Fig.~\ref{fig:1}(a).
As we show in the explicit calculations below, the low-energy excitations in the hole-doped system are localized exclusively in these AB and BA regions, effectively forming an ideal honeycomb lattice [see Fig.~\ref{fig:3}(a)], in agreement with previous work~\cite{pan23}. 
This localization gives rise to a spin-degenerate pair of graphene-like energy bands at the top of the noninteracting spectrum, indicated by the dashed lines in Fig.~\ref{fig:3}(d).
For a filling of $\nu = 2$ holes per moiré unit cell, the two topmost valence bands are half-filled, placing the Fermi level precisely at the Dirac band crossing located at the moiré Brillouin zone corners $\boldsymbol{\kappa}$ and $\boldsymbol{\kappa}'$.
While previous work~\cite{pan23} used an effective Hubbard-type model on the emergent honeycomb lattice, here we employ the realistic continuum model with long-range Coulomb interactions, which allows one to consistently account for the twist-angle dependence of the interaction strength and include contributions from remote bands.
The continuum model further allows us to systematically study further experimental tuning knobs, such as pressure, along with perturbations that are harder to control, such as strain.

\subsection{Uniaxial pressure}
\label{subsec:model-pressure}
A simple and effective way to tune the electronic structure of a moiré system is by applying uniaxial pressure along the out-of-plane direction.
This sensitivity arises from the strong dependence of the electronic spectrum on the interlayer spacing of the van-der-Waals-bonded TMD layers~\cite{brzezinska25}.
To model the effects of uniaxial pressure, we assume that the intralayer geometry remains unchanged.
We further simplify by taking the pressure-induced changes in the distance $d_{12}$ between inner and outer layers and the distance $d_{23}$ between the two inner layers to be equal,
$d_{12}(p) = d_{12}(0) + \Delta d_\perp(p)$, $d_{23}(p) = d_{23}(0) + \Delta d_\perp(p)$, 
where $p$ denotes the pressure difference from ambient conditions, $d_{12}(0)$ and $d_{23}(0)$ are the interlayer distances at ambient pressure, and $\Delta p_\perp$ is the pressure-induced change in interlayer spacing.
In this simplified model, the effect of uniaxial pressure amounts solely to a rescaling of a global interlayer tunneling scale.
The latter is expected to depend exponentially on $\Delta d_\perp$~\cite{laissardiere10}, motivating the modeling of the interlayer tunneling amplitudes as
\begin{align}
\begin{pmatrix}
V_{12}(p)\\ 
V_{23}^{(0)}(p) \\
V_{23}^{(1)}(p)
\end{pmatrix}
=
\rme^{- \beta \Delta d_\perp}
\begin{pmatrix}
V_{12}(0)\\ 
V_{23}^{(0)}(0) \\
V_{23}^{(1)}(0)
\end{pmatrix},
\end{align}
where $\beta$ is a material-dependent decay constant and $V_{12}(0)$, $V_{23}^{(0)}(0)$, and $V_{23}^{(1)}(0)$ correspond to the interlayer tunneling amplitudes at ambient pressure. 
Assuming the pressure-induced change in interlayer spacing is small, $\Delta d_\perp(p) \ll d_{12},d_{23}$, implies the linearized form
\begin{align}
\begin{pmatrix}
V_{12}(p)\\ 
V_{23}^{(0)}(p) \\
V_{23}^{(1)}(p)
\end{pmatrix}
=
\left(
1+ \frac{p}{p_0}
\right)
\begin{pmatrix}
V_{12}(0)\\ 
V_{23}^{(0)}(0) \\
V_{23}^{(1)}(0)
\end{pmatrix} + \mathcal O(p^2),
\end{align}
where $p_0$ is a material-dependent constant corresponding to the pressure required to double the interlayer tunneling amplitudes.
This simple model agrees reasonably well with ab-initio results for layered TMDs~\cite{brzezinska25}. From the comparison for the case of WSe$_2$, we estimate $p_0 \simeq 5\,\mathrm{GPa}$.
At the level of our qualitative analysis, we expect hydrostatic pressure to produce a similar effect, as it mainly impacts the interlayer spacing, while its influence on the intralayer geometry is expected to be less significant~\cite{du25}.

\subsection{Residual heterostrain}
\label{subsec:model-strain}
One of the most significant perturbations in moiré heterostructures is residual heterostrain, which corresponds to a relative in-plane distortion between adjacent layers. Even small amounts of heterostrain can substantially modify the moiré band structure and symmetry properties, and since it is generally expected to occur in realistic samples, it must be incorporated in any realistic analysis~\cite{xie19, parker21, kwan23}.
To illustrate the impact of strain on twisted double bilayer TMDs, we focus on uniaxial heterostrain applied between the two bilayers. Because the low-energy physics is dominated by the $\boldsymbol{\Gamma}$ valley, the quantitative effect of heterostrain on the electronic band structure is relatively small. Nonetheless, it explicitly breaks the symmetries that protect the linear band crossings at the $\boldsymbol{\kappa}$ points of the moir\'e Brillouin zone, with important consequences for the critical behavior at ultralow temperatures.

In $\mathbf K$-valley moiré heterostructures, strain generally has a twofold effect: it modifies the lattice geometry and shifts the microscopic $\mathbf K$ and $\mathbf K^\prime$ valleys due to anisotropic changes in the in-plane bond lengths, typically captured via minimal coupling to a phenomenological vector potential~\cite{pereira09,bi19,parker21}.
In contrast, an analogous shift of the $\boldsymbol{\Gamma}$ valley in the double bilayer TMDs studied here vanishes within the present continuum model, as it would violate the $C_{2y}$ symmetry of each individual layer, which forms the starting point of our analysis.
Consequently, we focus exclusively on the effects of strain-induced geometric changes. 
Importantly, while heterostrain preserves the $C_{2y}$ symmetry of the individual layers, it breaks this symmetry in the full continuum Hamiltonian by distorting the top and bottom bilayers differently.
The threefold rotation symmetry $C_{3z}$ about the out-of-plane $z$ axis is already broken at the single-layer level.
Effects beyond the present continuum model therefore do not introduce any additional symmetry breaking and are not expected to lead to further qualitative changes.

To incorporate heterostrain in the continuum model, we follow Ref.~\cite{bi19}, but unlike that work, we allow for generic (not necessarily small) twist angles $\theta$. Uniaxial strain can be characterized by two parameters: the strain magnitude $\epsilon$ and the strain direction $\phi$. Under strain, the in-plane position vectors are deformed as $\mathbf r \to \left(\mathbb{1}_2 + S(\epsilon, \phi)\right) \mathbf r$, with
\begin{align}
S(\epsilon, \phi) =  R(\phi)^{-1}
\begin{pmatrix}
\epsilon & 0 \\
0 & -\nu_{\mathrm{P}} \epsilon
\end{pmatrix}  R(\phi).
\end{align}
Here, $\nu_{\mathrm{P}}$ is the Poisson ratio of the TMD material, which we take as $\nu_\mathrm{P} = 0.2$ for WSe$_2$~\cite{kang13, zeng15}. The angle $\phi$ specifies the direction of strain relative to the $x$ axis, and $R(\phi)$ is the corresponding rotation matrix for a rotation by $\phi$ about the out-of-plane direction.
We consider a setup in which the top and bottom bilayers are first oppositely strained along the $x$ direction ($\phi = 0$), with strain magnitudes $-\epsilon/2$ on the top and $+\epsilon/2$ on the bottom, and are subsequently twisted relative to each other by $\mp \theta/2$ [cf. inset of Fig.~\ref{fig:6}(a)].
This corresponds to a transformation of the position vectors, $\mathbf r \to \mathcal M(\mp \theta/2, \mp \epsilon/2)\, \mathbf r$, with the minus sign for the top bilayer and the plus sign for the bottom bilayer, where the deformation matrix is defined as
\begin{align}
{M}(\theta, \epsilon) = {R}(\theta) \left(1 + S(\epsilon, 0)\right).
\end{align}
The reciprocal lattice vectors of the top and bottom bilayers transform as $\mathbf g_{i,\mathrm{t}/\mathrm{b}} = M(\mp \theta/2, \pm \epsilon/2)\, \mathbf g_i$, and the moir\'e reciprocal lattice vectors are given as $\mathbf G_i = \mathbf g_{i,\mathrm{t}} - \mathbf g_{i,\mathrm{b}}$.
Since the $\boldsymbol \Gamma$-valley dispersion in Eq.~\eqref{eq:gamma-dispersion} is written in terms of unstrained momenta, we introduce a modified dispersion
\begin{align}
h_{\boldsymbol \Gamma}(\mathbf q) = -\frac{\hbar^2}{2 m_{\boldsymbol \Gamma}}
\begin{pmatrix}
\left({M}(0, -\tfrac{\epsilon}{2})\, \mathbf q\right)^2 \mathbb 1_2 & 0 \\
0 & \left({M}(0, +\tfrac{\epsilon}{2})\, \mathbf q\right)^2 \mathbb 1_2
\end{pmatrix},
\end{align}
which effectively removes the strain from the wavevector $\mathbf q$ before evaluating the Hamiltonian.

\section{Hartree-Fock approach}
\label{sec:hartree-fock}

To investigate the phase diagram of the interacting model, we employ a Hartree-Fock mean-field decoupling, approximating the many-body ground state by a single Slater determinant that is optimized self-consistently. The Slater determinant is conveniently encoded in the one-body reduced density matrix, 
$P_{ns, n^\prime s^\prime}(\mathbf k) = \langle c^\dagger_{\mathbf k n^\prime s^\prime} c_{\mathbf k n s} \rangle$. 
The corresponding Hartree-Fock Hamiltonian takes the form~\cite{bultinck20, faulstich23}
\begin{align}
\mathcal{H_\mathrm{HF}}[P](\mathbf k) = h(\mathbf k) + h_\mathrm{H}[P](\mathbf k) + h_\mathrm{F}[P](\mathbf k)
\end{align}
with Hartree and Fock contributions
\begin{align}
h_{\text{H}}[P](\mathbf k) &= \sum_{\mathbf G} \frac{V_{\mathbf G}}{A} \Lambda(\mathbf k, \mathbf G) \sum_{\mathbf k^\prime} \operatorname{Tr}\left( P(\mathbf k^\prime) \Lambda(\mathbf k^\prime, \mathbf G)^\dagger \right),
\end{align}
and
\begin{align}
h_{\text{F}}[P](\mathbf k) &= - \sum_{\mathbf q} \frac{V_{\mathbf q}}{A} \Lambda(\mathbf k, \mathbf q) P(\mathbf k + \mathbf q) \Lambda(\mathbf k, \mathbf q)^\dagger,
\end{align}
respectively.
At zero temperature, the ground-state density matrix $P$ is determined by minimizing the Hartree-Fock energy functional
\begin{align} \label{eq:HF_energy}
E_{\mathrm{HF}}[P] = \sum_{\mathbf k} \Tr \left[ P \left(h(\mathbf k) + \frac{1}{2} h_{\mathrm{HF}}[P](\mathbf k) \right) \right],
\end{align}
where the factors of $1/2$ prevent double counting of the Coulomb interaction energy and we have introduced the shorthand $ h_{\mathrm{HF}}[P](\mathbf k) =  h_{\mathrm{H}}[P](\mathbf k) +  h_{\mathrm{F}}[P](\mathbf k)$. The minimization is carried out using the optimal damping algorithm~\cite{cances00}.

\section{Strong-coupling limit}
\label{sec:strong-coupling}

In the following, we investigate possible correlated orders at small twist angles $\theta \sim 1^\circ$, where interactions dominate the competition between the different orders. Our approach parallels the strong-coupling treatment of twisted bilayer graphene~\cite{bultinck20}: we restrict the analysis to the two spin-degenerate graphene-like bands, introduce a sublattice basis for this subspace, and fix convenient representations of the symmetry operators. These symmetries are then used to derive a simplified form of the overlap matrices $\Lambda(\mathbf k, \mathbf q)$ in the sublattice basis, which enables explicit evaluation of the competing states' energies in the flat-band limit. We summarize the main steps here, with further details provided in Appendix~\ref{app:strong-coupling}.

By analogy with the strong-coupling limit of twisted bilayer graphene, we begin by assuming an approximate chiral symmetry $\mathcal C$ in the noninteracting spectrum. We demonstrate in Appendix~\ref{app:energy-scales} that this assumption is justified at small twist angles near $1^\circ$, where the bandwidth of the two spin-degenerate low-energy bands is narrow [cf.~Fig.~\ref{fig:3}(f)], even though no parameter choice of the continuum model renders the symmetry exact. The chiral limit nevertheless provides a useful framework for building intuition about the competing orders at small twist angles. Antichiral terms present in the realistic model are then included as perturbative corrections.

Although the individual TMD layers lack sublattice-exchange symmetry on the microscopic honeycomb lattice, the emergent honeycomb lattice at hole doping $\nu = 2$ exhibits an effective $C_{2z}$ lattice rotation symmetry on the moiré scale. 
This arises because the AB and BA regions of the microscopic lattice, shown in Fig.~\ref{fig:1}(a), are related by the microscopic $C_{2y}$ symmetry. As a result, observables such as the charge density on the two sites of the emergent honeycomb unit cell are equal unless the ground state breaks $C_{2y}$ symmetry.

At each wavevector, we restrict the full Hilbert space to the four-dimensional subspace corresponding to the two spin-degenerate low-energy bands. In this subspace, we adopt a sublattice basis, in which the two-fold-lattice-rotation, time-reversal, and chiral symmetries are represented by
\begin{align}
C_{2z} & = \sigma_x , &
\mathcal{T} & = \rmi s_y \mathcal K, &
\mathcal{C}& = \sigma_z,
\end{align}
where
$\mathcal K$ denotes complex conjugation, $\sigma_x$ and $\sigma_z$ are Pauli matrices acting on the sublattice degrees of freedom, and the Pauli matrix $s_y$ acts on the spin degree of freedom.
Note that $C_{2z}$ and $\mathcal T$ invert wavevectors, whereas $\mathcal C$ leaves them unchanged.
Decomposing the overlap matrices into symmetric and antisymmetric parts under chiral transformations, $\Lambda(\mathbf k, \mathbf q) = \Lambda_{\mathrm S}(\mathbf k, \mathbf q) + \Lambda_{\mathrm A}(\mathbf k, \mathbf q)$, with $\Lambda_{\mathrm{S} / \mathrm{A}}(\mathbf k, \mathbf q) = \frac{1}{2} \left( \Lambda(\mathbf k, \mathbf q) \pm \sigma_z \Lambda(\mathbf k, \mathbf q) \sigma_z \right)$ enforces the forms
\begin{align}
\Lambda_{\mathrm S}(\mathbf k, \mathbf q) & = F_{\mathrm S}(\mathbf k, \mathbf q) \rme^{\rmi \Phi_{\mathrm S}(\mathbf k, \mathbf q) \sigma_z},
\\
\Lambda_{\mathrm A}(\mathbf k, \mathbf q) & = F_{\mathrm A}(\mathbf k, \mathbf q) \sigma_x \rme^{\rmi \Phi_{\mathrm A}(\mathbf k, \mathbf q) \sigma_z},
\end{align}
diagonal in spin indices, 
where $F_{\mathrm S / \mathrm A}$ and $\Phi_{\mathrm S / \mathrm A}$ are real-valued functions.

To facilitate further analytical progress, we rewrite the interacting Hamiltonian for the low-energy bands as~\cite{bultinck20}
\begin{align} \label{eq:strong-coupling-model}
\mathcal H &= \sum_{\mathbf k} c^\dagger_{\mathbf k} \tilde h(\mathbf k) c_{\mathbf k} + \frac{1}{2A} \sum_{\mathbf q} V_{\mathbf q} \delta \rho_{\mathbf q} \delta \rho_{-\mathbf q} + \mathrm{const.},
\end{align}
where
$\delta \rho_{\mathbf q} = \rho_{\mathbf q} - \bar \rho_{\mathbf q}$ 
is the charge density relative to the average
$\bar \rho_{\mathbf q} = \frac{1}{2} \sum_{\mathbf G, \mathbf k} \delta_{\mathbf G, \mathbf q} \Tr \Lambda (\mathbf k, \mathbf G)$.  
The renormalized dispersion $\tilde h(\mathbf k)$ is defined in Appendix~\ref{app:strong-coupling}.
Neglecting both the dispersion $\tilde h$ and the antichiral contribution $\Lambda_{\mathrm A}(\mathbf k, \mathbf q)$ yields the chiral strong-coupling Hamiltonian
\begin{align}
\mathcal H_{\mathrm S} = \frac{1}{2A} \sum_{\mathbf q} V_{\mathbf q} \delta \rho_{\mathbf q}^{\mathrm S} \delta \rho_{-\mathbf q}^{\mathrm S},
\end{align}
where $\delta \rho_{\mathbf q}^{\mathrm S}$ denotes the charge density in the chiral limit, corresponding to $\Lambda_{\mathrm A}(\mathbf k, \mathbf q) \to 0$.
Note that $\mathcal{H}_{\mathrm S}$ is positive semidefinite, so its ground-state manifold consists precisely of all many-body states annihilated by it.
The antichiral contribution to the strong-coupling Hamiltonian is
\begin{align}
\mathcal H_{\mathrm A} = \frac{1}{2A} \sum_{\mathbf q} V_{\mathbf q} (\delta \rho_{\mathbf q}\delta \rho_{-\mathbf q} - \delta \rho_{\mathbf q}^{\mathrm S}\delta \rho_{-\mathbf q}^{\mathrm S}),
\end{align}
and is expected to act as a small perturbation at twist angles $\theta$ near $1^\circ$.

For calculations in the strong-coupling limit, Slater determinant states $P$ are conveniently parametrized as $P = \frac{1}{2}(\mathbb{1}_4 + Q)$ in the sublattice basis, with $Q^\dagger = Q$, $Q^2 = \mathbb{1}_4$, and $\Tr Q = 0$ at half filling of the graphene-like bands, corresponding to a hole doping density $\nu = 2$ per moiré unit cell. Motivated by our numerical results in Sec.~\ref{sec:phase-diagram} and by the analogy with the strong-coupling limit of twisted bilayer graphene~\cite{bultinck20}, we focus specifically on the energy competition between Slater determinant states satisfying $[Q, \sigma_z] = 0$, given by
\begin{align} \label{eq:candidate-states}
\text{Charge-density-wave state}&: \quad Q = \sigma_z s_0, \\
\text{Ferromagnetic state}&: \quad Q = \mathbf n \cdot \mathbf s, \\
\text{Antiferromagnetic state}&: \quad Q = \sigma_z \mathbf n \cdot \mathbf s.
\end{align}
Here, $\mathbf{s} = (s_x, s_y, s_z)$ are Pauli matrices acting in spin space, $s_0$ is the identity matrix in spin space, and $\mathbf{n}$ is an arbitrary unit vector specifying the direction of the (staggered) magnetization.
The charge-density-wave state breaks the sublattice symmetry of the emergent honeycomb lattice, while the ferromagnetic state breaks both SU(2) spin-rotation and time-reversal symmetries. The antiferromagnetic state breaks SU(2) spin-rotation, time-reversal, and sublattice symmetries, but preserves the combination of time reversal and sublattice exchange.
The sublattice symmetry breaking on the emergent honeycomb lattice corresponds to a $C_{2y}$ symmetry breaking of the microscopic moiré system.
All three candidate ground states are Slater determinants, so their internal energy $E_{\mathrm{HF}, \mathrm S}[P]$ with respect to $\mathcal H_{\mathrm S}$ can be evaluated at the mean-field level, yielding
\begin{align}
E_{\mathrm{HF}, \mathrm S}[P_{\text{FM}/\text{AFM}}] = 0
\end{align}
for the ferromagnetic and antiferromagnetic states, and
\begin{align} \label{eq:energy-cdw}
E_{\mathrm{HF}, \mathrm S}[P_\text{CDW}] = \frac{2}{A} \sum_{\mathbf{G}} V_{\mathbf{G}} \left[ \sum_{\mathbf k} F_{\mathrm S}(\mathbf k, \mathbf G) \sin\left(\Phi_{\mathrm S}(\mathbf k, \mathbf G)\right) \right]^2
\end{align}
for the charge-density-wave state. Since no symmetry enforces the vanishing of the inner sum in Eq.~\eqref{eq:energy-cdw}, only the two magnetic orders generically emerge as ground states of the chiral strong-coupling Hamiltonian $\mathcal H_{\mathrm S}$.

\begin{figure}[tb!]
\includegraphics[width=\linewidth]{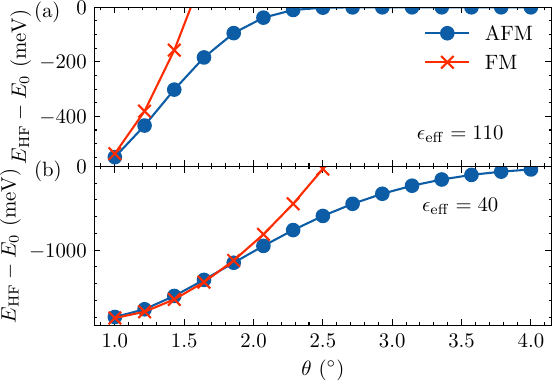}
\caption{
(a)~Hartree-Fock energy $E_\mathrm{HF}$ of the antiferromagnetic (blue) and ferromagnetic (red) states relative to the noninteracting ground-state energy $E_0$ in the full model [Eq.~\eqref{eq:full-model}] for an effective permittivity $\epsilon_\mathrm{eff} = 110$, representative of realistic values for twisted double bilayer WSe$_2$. Calculations were performed on an $18 \times 18$ momentum-space grid, keeping two bands per spin species, with initial states preserving the respective symmetries of each candidate phase. At small twist angles, the two states are nearly degenerate, reflecting the approximate chiral symmetry, yet the antiferromagnetic state is ultimately favored across the entire range.
(b)~Same as (a) but for $\epsilon_\mathrm{eff} = 40$, enhancing the relative weight of the antichiral interaction compared with the kinetic contribution. In this case, the ferromagnetic state is favored at sufficiently small twist angles.
}
\label{fig:2}
\end{figure}

Because the antiferromagnetic and ferromagnetic states are degenerate in the chiral limit, it is essential to account for the previously neglected perturbations: the antichiral interaction and the kinetic contribution from the band dispersion.
A calculation of the energy contribution $E_{\mathrm{HF}, \mathrm A}$ arising from the antichiral part $\mathcal H_{\mathrm A}$ of the strong-coupling Hamiltonian gives
\begin{align}
E_{\mathrm{HF}, \mathrm A}[P_\text{FM}] = 0
\end{align}
for the ferromagnetic state and
\begin{align} \label{eq:energy-afm}
E_{\mathrm{HF}, \mathrm A}[P_\text{AFM}] = \frac{1}{A} \sum_{\mathbf q, \mathbf k} V_{\mathbf q} [F_{\mathrm A}(\mathbf k, \mathbf q)]^2
\end{align}
for the antiferromagnetic state, with the latter being strictly positive if there are antichiral contributions to the overlap matrices.
Therefore, when the antichiral interaction is the dominant perturbation, the ferromagnetic state is energetically preferred.
Similarly, we show in Appendix~\ref{app:effect-of-dispersion} that the dispersion favors the antiferromagnetic state.

To illustrate the competition between the two magnetic orders in the presence of both antichiral and kinetic contributions, Fig.~\ref{fig:2} shows numerical results from a Hartree-Fock analysis of the full interacting continuum model as a function of twist angle $\theta$ for two different effective permittivities, $\epsilon_\text{eff} = 110$ and $\epsilon_\text{eff} = 40$.
The relative energy difference between the two magnetic states decreases with decreasing twist angle, reflecting the approximate chiral symmetry at small angles. For the larger permittivity $\epsilon_\text{eff} = 110$, which is of the order of realistic values, the antiferromagnetic state is favored because the kinetic contribution dominates over the antichiral terms, as shown in Fig.~\ref{fig:2}(a).
By contrast, reducing the effective permittivity enhances the relative strength of the antichiral interaction, stabilizing the ferromagnetic state over a range of twist angles, as illustrated for $\epsilon_\text{eff} = 40$ in Fig.~\ref{fig:2}(b).
This interplay between antichiral interactions and the kinetic contribution underlies the appearance of both ferromagnetic and antiferromagnetic regions in the phase diagram discussed in the following section.

\section{Angle-tuned Phase diagram}
\label{sec:phase-diagram}

\begin{figure*}[tb!]
\includegraphics[width=\linewidth]{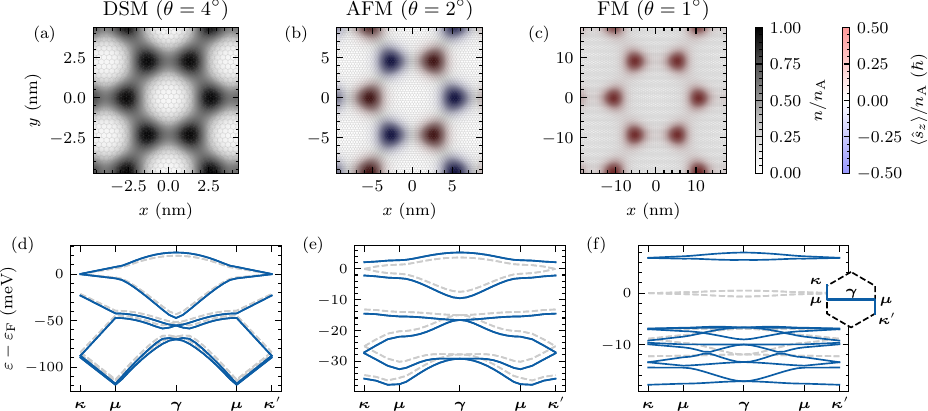}
\caption{
(a)~Charge density $n(\mathbf R)$ (grayscale) and spin density $\langle \hat{s}_z \rangle(\mathbf R)$ (overlaid color scale, opacity indicating magnitude) in the Dirac semimetal (DSM) ground state of the full Hamiltonian [Eq.~\eqref{eq:full-model}] at twist angle $\theta = 4^\circ$ and effective permittivity $\epsilon_\text{eff} = 110$, representative of realistic values for twisted double bilayer WSe$_2$. Light gray lines indicate the microscopic lattices of the inner TMD layers. The charge density forms an emergent honeycomb pattern on the moiré scale. Densities are normalized to the value $n_{\mathrm A}$ at the A site of the emergent honeycomb lattice.
In the Dirac semimetal phase, the spin density vanishes, rendering the overlaid color scale transparent.
(b)~Same as~(a), but for twist angle $\theta = 2^\circ$.
The ground state is antiferromagnetic (AFM), with antiparallel spins on the two sites of the emergent honeycomb lattice. The moiré scale is enlarged compared to (a).
(c)~Same as~(a), but for twist angle $\theta = 1^\circ$ and reduced effective permittivity $\epsilon_\text{eff} = 40$. The ground state is ferromagnetic (FM), with parallel spins on the two sites of the emergent honeycomb lattice.
(d)~Electronic spectrum at twist angle $\theta = 4^\circ$ and effective permittivity $\epsilon_\text{eff} = 110$, representative of realistic values for twisted double bilayer WSe$_2$. Dirac points at the moiré Brillouin zone corners $\boldsymbol{\kappa}$ and $\boldsymbol{\kappa}’$ remain gapless and spin degenerate. Dashed gray lines show the noninteracting bands for comparison. Energies are referenced to the Fermi level $\varepsilon_\mathrm{F}$ at filling $\nu = 2$ holes per moiré unit cell.
(e)~Same as (d), but for twist angle $\theta = 2^\circ$. In contrast to the Dirac semimetal, the spectrum is fully gapped while remaining spin-degenerate, realizing an antiferromagnetic insulator.
(f)~Same as (d), but for twist angle $\theta = 1^\circ$ and reduced effective permittivity $\epsilon_\text{eff} = 40$.
The spectrum is fully gapped and spin-split, realizing a ferromagnetic insulator.
The inset indicates the path through the moiré Brillouin zone along which the energy bands are shown.
}
\label{fig:3}
\end{figure*}

To explore the quantum phase diagram of the full interacting Hamiltonian as a function of twist angle, we perform self-consistent Hartree-Fock calculations on an $18\times18$ momentum-space grid, truncating the Hamiltonian to six bands per spin species.
Consistent with expectations from the strong-coupling analysis, we identify three distinct types of ground states, depending on the twist angle $\theta$ and effective permittivity $\epsilon_\text{eff}$:
(i)~A fully symmetric Dirac semimetal;
(ii)~A ferromagnetic state breaking $\mathrm{SU}(2)$ spin rotation and time-reversal symmetry; and
(iii)~An antiferromagnetic state breaking $\mathrm{SU}(2)$ spin rotation, time-reversal, and sublattice symmetry of the emergent honeycomb lattice, while preserving the combined operation of time reversal and sublattice exchange.

Figures~\ref{fig:3}(a-c) display the real-space charge density
$n(\mathbf R) = \sum_{\ell, s} \langle f^\dagger_{\mathbf R, \ell, s} f_{\mathbf R, \ell, s} \rangle$
and out-of-plane spin expectation value
$\langle \hat s_z \rangle(\mathbf R) = \frac{\hbar}{2} \sum_{\ell,s,s’} \langle f^\dagger_{\mathbf R, \ell, s} (s_z)_{s s’} f_{\mathbf R, \ell, s’} \rangle$
of the ground state for twist angles $4^\circ$, $2^\circ$, and $1^\circ$, at a fixed effective permittivity $\epsilon_\text{eff} = 110$, representative for realistic values for twisted double bilayer WSe$_2$.
For clarity, the Hartree-Fock calculations are initialized so that any magnetic ordering occurs only along the out-of-plane direction. The corresponding electronic band structures are shown in Figs.~\ref{fig:3}(d-f).
At $\theta = 4^\circ$ [Figs.~\ref{fig:3}(a,d)], the system realizes a Dirac semimetal ground state, with spin-degenerate bands that are gapless up to finite-size effects. The low-energy spectrum can be viewed as a renormalized version of the noninteracting graphene-like bands, featuring two spin-degenerate Dirac cones at the corners $\boldsymbol \kappa$ and $\boldsymbol \kappa^\prime$ of the moiré Brillouin zone.
The real-space charge density [Fig.~\ref{fig:3}(a)] shows localization of low-energy excitations in the AB and BA regions of the moiré lattice, revealing the emergent honeycomb lattice with preserved sublattice symmetry. Integrating $n(\mathbf R)$ over the moiré unit cell yields a total of two, consistent with a filling of $\nu = 2$ holes per moiré unit cell.
Reducing the twist angle narrows the noninteracting bandwidth, enhancing interaction effects and destabilizing the Dirac semimetal.
At $\theta = 2^\circ$ [Figs.~\ref{fig:3}(b,e)], interactions are strong enough to stabilize an antiferromagnetic phase. In this state, spontaneous breaking of SU(2) spin-rotation symmetry opens a spectral gap $\Delta = 4.3\,\mathrm{meV}$, and the real-space spin density [Fig.~\ref{fig:3}(b)] shows antiparallel spins on the two sublattices of the emergent honeycomb lattice, corresponding to Néel-type antiferromagnetic order on the moiré scale.
Reducing the twist angle brings the system closer to the chiral limit, decreasing the energy difference between the antiferromagnetic and competing ferromagnetic states. 
At $\theta = 1^\circ$, a ferromagnetic insulator emerges at sufficiently small effective permittivity, as shown for $\epsilon_\text{eff} = 40$ in Figs.~\ref{fig:3}(c,f). The state exhibits spin-split bands and a finite net magnetization across the moiré unit cell.

\begin{figure}[tb!]
\includegraphics[width=\linewidth]{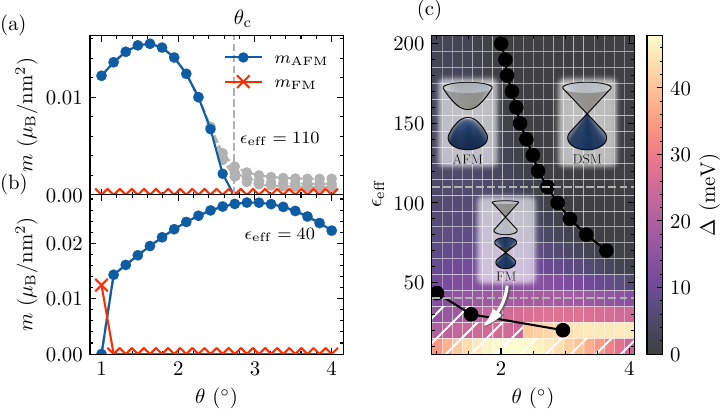}
\caption{
(a)~Total and staggered magnetization densities $m_\mathrm{FM}$ and $m_\mathrm{AFM}$ at the centers of the AB and BA regions of the moiré lattice as functions of twist angle $\theta$ for fixed effective permittivity $\epsilon_\text{eff} = 110$, representative of realistic values for twisted double bilayer WSe$_2$. Gray points show $m_\mathrm{AFM}$ at finite system sizes $L = 12, 15, 18$, while blue points indicate the extrapolation to the thermodynamic limit. The results reveal a continuous quantum phase transition from an antiferromagnetic insulator for $\theta < \theta_\mathrm{c}$ to a symmetric Dirac semimetal for $\theta > \theta_\mathrm{c}$, with critical twist angle $\theta_\mathrm{c} \simeq 2.7^\circ$.
(b)~Same as~(a), but for reduced effective permittivity $\epsilon_\text{eff} = 40$. At small twist angles, antiferromagnetic and ferromagnetic orders compete, leading to a strong first-order transition through a level crossing near $\theta \simeq 1.1^\circ$. Below this angle, the staggered magnetization vanishes and the ground state develops a finite net magnetization.
(c)~Quantum phase diagram as a function of twist angle $\theta$ and effective permittivity $\epsilon_\text{eff}$. Regions of antiferromagnetic (AFM) order, ferromagnetic (FM) order, and the symmetric Dirac semimetal (DSM) phase are identified. Insets indicate the low-energy band structure. The antiferromagnetic-to-ferromagnet transition is first-order, while the Dirac-semimetal-to-antiferromagnetic-insulator transition is continuous and belongs to the relativistic Gross-Neveu-Heisenberg universality class. Dashed gray lines indicate parameter cuts shown in (a) and (b).
}
\label{fig:4}
\end{figure}

To track the evolution of the magnetic orders, we define the total and staggered magnetization densities at the centers of the AB and BA regions of the moiré lattice as
\begin{align}
m_{\mathrm{FM} / \mathrm{AFM}} &= \sum_\ell \left | \langle \boldsymbol \mu_{\mathbf R_{\mathrm A}, \ell} \rangle \pm \langle \boldsymbol \mu_{\mathbf R_{\mathrm B}, \ell} \rangle \right |,
\end{align}
where
\begin{align}
\langle \boldsymbol \mu_{\mathbf R, \ell} \rangle = \Tr \left[- \mu_{\mathrm B} \mathbf s 
\begin{pmatrix}
\langle f^\dagger_{\mathbf R, \ell, \uparrow} f_{\mathbf R, \ell, \uparrow} \rangle & \langle f^\dagger_{\mathbf R, \ell, \downarrow} f_{\mathbf R, \ell, \uparrow} \rangle \\
\langle f^\dagger_{\mathbf R, \ell, \uparrow} f_{\mathbf R, \ell, \downarrow} \rangle & \langle f^\dagger_{\mathbf R, \ell, \downarrow} f_{\mathbf R, \ell, \downarrow} \rangle
\end{pmatrix}
\right]
\end{align}
represents the magnetization density in layer $\ell$ at position $\mathbf R$, $\mathbf R_{\mathrm A}$ and $\mathbf R_{\mathrm B}$ denote the positions of the AB and BA regions of the moiré lattice, and $\mu_{\mathrm B}$ is the Bohr magneton.

Figure~\ref{fig:4}(a) shows the total and staggered magnetization densities $m_\mathrm{FM}$ and $m_\mathrm{AFM}$ as functions of twist angle $\theta$ for a fixed effective permittivity $\epsilon_\text{eff} = 110$, representative of realistic values for twisted double bilayer WSe$_2$.
The total magnetization vanishes across the range of twist angles studied, while the staggered magnetization remains finite at small twist angles. For very small twist angles around $1^\circ$, $m_\mathrm{AFM}$ initially increases with increasing $\theta$, reaching a maximum near $\sim 1.6^\circ$, and then decreasing rapidly toward zero.
Extrapolating the finite-size results (gray data points) to the thermodynamic limit $1/L \to 0$ (blue data points) indicates a continuous transition from the antiferromagnetic insulator at $\theta < \theta_\mathrm{c}$ to the symmetric Dirac semimetal at $\theta > \theta_\mathrm{c}$, with a critical twist angle $\theta_\mathrm{c} \simeq 2.7^\circ$.
As a cross-check, we have performed a crossing-point analysis following Ref.~\cite{biedermann25}, using the renormalization-group invariants $R_{\Delta} = L \Delta / \Delta_0$ (with $\Delta_0 \simeq 1.6\,\mathrm{eV}$, the direct band gap of monolayer WSe$_2$~\cite{yun12}) and $R_m = L m_{\mathrm{AFM}} a_0^2 / \mu_{\mathrm B}$, constructed from the interaction-induced gap $\Delta$ and staggered magnetization $m_\mathrm{AFM}$. The curves for $R_\Delta$ and $R_m$ as functions of $\theta$ at different system sizes (not shown) reveal a unique crossing at $\theta_\mathrm{c} \simeq 2.7^\circ$, confirming the continuous quantum phase transition between the antiferromagnetic insulator and the symmetric Dirac semimetal.

Figure~\ref{fig:4}(b) shows the total and staggered magnetization densities as functions of twist angle $\theta$ for a smaller effective permittivity $\epsilon_\text{eff} = 40$. In this case, the maximum of the staggered magnetization shifts to larger twist angles, and the transition between the antiferromagnet and the Dirac semimetal occurs at a critical angle beyond the range considered here. At small twist angles, the antiferromagnetic order competes with ferromagnetic order, giving rise to a strong first-order transition via a level crossing around $\theta \simeq 1.1^\circ$. Below this angle, the staggered magnetization vanishes and the ground state exhibits a finite net magnetization.

Figure~\ref{fig:4}(c) presents the quantum phase diagram as a function of twist angle $\theta$ and effective permittivity $\epsilon_\text{eff}$.
At large twist angles and realistic $\epsilon_\text{eff}$, the Dirac semimetal is stable, while decreasing $\theta$ drives a continuous transition into the antiferromagnetic insulator. The semimetal-to-insulator phase boundary in Fig.~\ref{fig:4}(c) is obtained from a crossing-point analysis of the renormalization group invariant $R_m$, consistent with the order-parameter extrapolation in Fig.~\ref{fig:4}(a).
At very small angles ($\sim 1^\circ$) and reduced $\epsilon_\text{eff}$, a ferromagnetic insulator emerges. The discontinuous antiferromagnet-to-ferromagnet boundary is determined from linear extrapolation of the energy offset $\Delta E = E_{\mathrm{HF}}[P] - E_{\mathrm{HF}}[P_0]$ across the transition, consistent with the level crossing in Fig.~\ref{fig:4}(b).

We reiterate that for comparison with experiments at small twist angles, values of $\epsilon_\text{eff} \sim \mathcal O(100)$ are realistic~\cite{goodwin19, biedermann25, munoz25}.
In Appendix~\ref{app:permittivity}, we estimate $\epsilon_\text{eff}$ in the Dirac semimetal phase using the random phase approximation~\cite{pizarro19}.
For the range of twist angles $\theta$ considered here, $\epsilon_\text{eff}$ is dominated by internal screening and therefore exhibits a strong dependence on $\theta$.
In the Dirac semimetal phase, $\epsilon_\text{eff}(\theta)$ increases upon reducing $\theta$, as the narrowing of the low-energy bands enhances electronic screening.
For twisted double bilayer WSe$_2$ at $\theta \simeq 2.7^\circ$, corresponding to the experimentally observed critical angle $\theta_\mathrm{c}$, we obtain $\epsilon_\text{eff}(2.7^\circ) \simeq 82$, assuming a substrate permittivity of $\epsilon_\text{env} = 5$.
This estimate is in reasonable agreement with the semimetal-to-insulator phase boundary shown in Fig.~\ref{fig:4}(c), which occurs at a critical permittivity $\epsilon_{\text{eff},\mathrm{c}} \simeq 112$ for $\theta = 2.7^\circ$.
In the gapped phase, $\epsilon_\text{eff}(\theta)$ is expected to decrease upon reducing $\theta$, reflecting the opening of an excitation gap. Its magnitude can be inferred by matching the extrapolated Hartree-Fock gap $\Delta$ to the experimental data from Ref.~\cite{ma24}, which yields $\epsilon_\text{eff}(2.5^\circ) \simeq 109$ and $\epsilon_\text{eff}(1.9^\circ) \simeq 51$, again in reasonable agreement with the expectation based on the random-phase-approximation estimates in the gapless phase.
This suggests that the ferromagnetic phase may also be realized in twisted double bilayer WSe$_2$ at small twist angles $\theta \lesssim 1^\circ$.

\section{Pressure-tuned phase diagram}
\label{sec:pressure}

To demonstrate the tunability of twisted double bilayer TMDs, we show that the Dirac-semimetal-to-antiferromagnetic-insulator transition can also be induced by applying uniaxial out-of-plane pressure, using the model introduced in Sec.~\ref{subsec:model-pressure}.
Uniaxial pressure enhances interlayer tunneling, reducing the noninteracting bandwidth and thereby amplifying interaction effects. For samples with twist angles just above the ambient-pressure phase boundary, uniaxial pressure can drive a transition from the symmetric Dirac semimetal to the symmetry-broken antiferromagnetic insulator.

\begin{figure}[tb!]
\includegraphics[width=\linewidth]{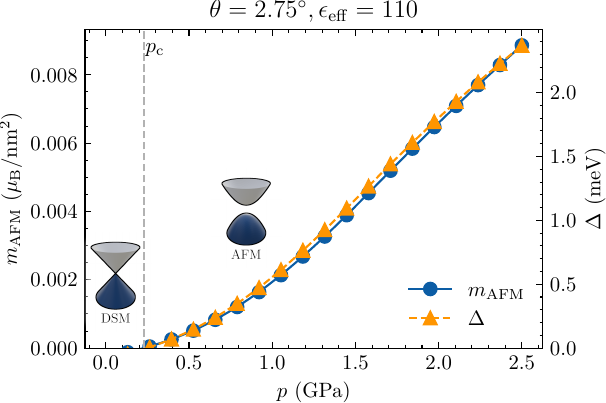}
\caption{%
Staggered magnetization density $m_\mathrm{AFM}$ (blue) and interaction-induced gap $\Delta$ (orange) as functions of applied uniaxial pressure $p$ along the out-of-plane direction, for fixed twist angle $\theta = 2.75^\circ$ and effective permittivity $\epsilon_\mathrm{eff} = 110$.
At low pressures, the ground state is a fully symmetric Dirac semimetal. Increasing pressure drives the system into an antiferromagnetic insulator. The continuous transition occurs at a critical pressure $p_\mathrm{c} \simeq 0.2\,\mathrm{GPa}$, with its precise value strongly dependent on the sample's twist angle.
}
\label{fig:5}
\end{figure}

Figure~\ref{fig:5} illustrates this effect: the staggered magnetization density $m_\text{AFM}$ and the spectral gap $\Delta$, extrapolated to the thermodynamic limit, are shown as functions of pressure $p$ for a sample with twist angle $\theta = 2.75^\circ$, just above the ambient-pressure critical angle $\theta_\mathrm{c}(0) = 2.7^\circ$. The results reveal a continuous transition from the Dirac semimetal ($p < p_\mathrm{c}$) to the antiferromagnetic insulator ($p > p_\mathrm{c}$), with a critical pressure $p_\mathrm{c} \simeq 0.2\,\mathrm{GPa}$ at $\theta - \theta_\mathrm{c}(0) = 0.05^\circ$. Consistent values of $p_\mathrm{c}$ are obtained from crossing-point analyses of the renormalization group invariants $R_m$ and $R_\Delta$ (not shown).

The critical pressure depends sensitively on the twist angle. For example, if the offset from the ambient-pressure phase boundary doubles to $\theta - \theta_\mathrm{c}(0) = 0.1^\circ$, the required pressure increases to $p_\mathrm{c} \simeq 0.6\,\mathrm{GPa}$, see Appendix~\ref{app:pressure}.
This strong dependence highlights uniaxial pressure as a highly effective tuning knob for accessing and traversing the critical region above the Dirac-semimetal-to-antiferromagnetic-insulator quantum critical point.

\section{Strain effects}
\label{sec:strain}

For comparison with experiments, it is essential to account for the effects of finite heterostrain~\cite{xie19, parker21, kwan23}.
We focus on uniaxial strain applied along the in-plane $x$ direction, modeled as described in Sec.~\ref{subsec:model-strain}.
Finite strain explicitly breaks the microscopic $C_{2y}$ symmetry, distorting the emergent honeycomb lattice and lifting its $C_{2z}$ rotation symmetry. Consequently, the Dirac cones acquire a gap even in the noninteracting limit, and at large twist angles the interacting ground state becomes a nematic band insulator without spontaneous symmetry breaking, with the single-particle gap $\Delta_\epsilon$ determined by the strain magnitude.

\begin{figure}[tb!]
\includegraphics[width=\linewidth]{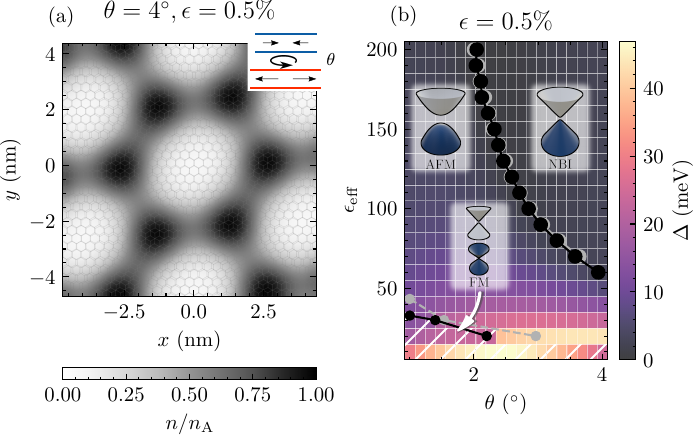}
\caption{
(a)~Charge density $n(\mathbf R)$ at twist angle $\theta = 4^\circ$ and effective permittivity $\epsilon_\text{eff} = 110$ in the presence of uniaxial heterostrain $\epsilon = 0.5\%$ along the $x$ direction. The inset illustrates the sample geometry: before introducing the twist $\theta$, the top (bottom) bilayer is compressed (stretched) along $x$ by $\epsilon/2$ of its length.
(b)~Quantum phase diagram as a function of twist angle $\theta$ and effective permittivity $\epsilon_\text{eff}$ in the presence of uniaxial heterostrain $\epsilon = 0.5\%$ along the $x$ direction. Gray markers indicate the phase boundaries without strain. At large twist angles, the ground state is a nematic band insulator (NBI) with a small strain-induced gap.
}
\label{fig:6}
\end{figure}

Figure~\ref{fig:6}(a) shows the distorted real-space charge density induced by heterostrain of magnitude $\epsilon = 0.5\%$ at a twist angle of $\theta = 4^\circ$.
For this strain strength, the strain-induced gap is small compared to the low-energy bandwidth, $\Delta_\epsilon \sim 1\,\mathrm{meV}$, so the system will still appear semimetallic at temperatures above $\Delta_\epsilon/k_\mathrm{B} \sim 10\,\mathrm{K}$.
Reducing the twist angle in the presence of finite heterostrain drives a transition from the band insulator to an antiferromagnetic insulator that spontaneously breaks SU(2) spin symmetry.
At sufficiently small effective permittivity $\epsilon_\text{eff}$, further decreasing the twist angle stabilizes a ferromagnetic insulator.

Figure~\ref{fig:6}(b) shows the quantum phase diagram as a function of twist angle $\theta$ and effective permittivity $\epsilon_\text{eff}$ in the presence of heterostrain with magnitude $\epsilon = 0.5\%$.
For comparison, the corresponding phase boundaries in the unstrained case $\epsilon = 0\%$ [Fig.~\ref{fig:4}(c)] are shown in gray.
The comparison reveals that strain leads only to minor shifts of the phase boundaries.
The staggered magnetization density $m_\text{AFM}$ again decreases continuously with increasing $\theta$, indicating a continuous transition from the antiferromagnetic insulator to the band insulator.
We note that in the strained case, $m_\text{AFM}$ already vanishes on finite-size systems in the band-insulating phase due to the broken $C_{2y}$ symmetry.
Consequently, the critical twist angles $\theta_\mathrm{c}$ shown in Fig.~\ref{fig:6}(b) are determined solely from the antiferromagnetic side, using a linear extrapolation of $m_\text{AFM}$ towards zero.

While the strain magnitudes considered here only weakly affect the phase boundaries, we show in the following section that they strongly influence the critical behavior in the quantum critical regime above the antiferromagnetic-insulator-to-band-insulator transition.

\section{Quantum critical behavior}
\label{sec:critical}

We now turn to the universal critical behavior expected when the antiferromagnetic order is suppressed by increasing the twist angle or reducing the applied pressure.
We first consider the ideal case of strain-free samples, and then discuss the impact of finite heterostrain present in realistic devices.

\subsection{Strain-free samples}

The Hartree-Fock analysis in Secs.~\ref{sec:phase-diagram} and \ref{sec:pressure} suggested a continuous quantum phase transition between the symmetric Dirac semimetal and the antiferromagnetic insulator, driven by twist angle $\theta$ or uniaxial pressure $p$.
Near the phase transition, however, fluctuations become important and Hartree-Fock theory may no longer be adequate.
To accurately describe this physics, we derive the most general effective low-energy field theory consistent with the microscopic symmetries of the model and the nature of the adjacent phases, suitable for further analysis within a renormalization group approach.
%
%
This quantum field theory can be deduced from symmetry considerations:
Define the eight-component Dirac fermion field
$\psi_\mathbf{k} = (c_{\boldsymbol{\kappa} + \mathbf k,n,s}, c_{\boldsymbol{\kappa}' + \mathbf k,n,s})$,
at long wavelengths $\mathbf k \ll \pi/a_\mathrm{M}$, where $a_\mathrm{M} \simeq a_0 / \theta$ denotes the moiré lattice scale and $c_{\boldsymbol{\kappa} + \mathbf k,n,s}$ ($c_{\boldsymbol{\kappa}' + \mathbf k,n,s}$) annihilates an electron at wavevector $\mathbf k$ relative to the moiré minivalley $\boldsymbol{\kappa}$ ($\boldsymbol{\kappa}’$) with band index $n=1,2$ and spin $s = {\uparrow,\downarrow}$.
We introduce the antiferromagnetic order parameter $\vec \varphi$, constructed from the long-wavelength staggered magnetization.
Under SU(2) spin rotations, $\vec \varphi$ transforms as a vector.
Consequently, the most general symmetry-allowed low-energy effective Lagrangian takes the form
\begin{multline} \label{eq:field-theory}
\mathcal L = 
\psi^\dagger (\partial_\tau + v_\mathrm{F} \gamma_0\vec\gamma \cdot \nabla) \psi 
+ \frac{1}{2} \vec \varphi \cdot (r - \partial_\tau^2 - v_\mathrm{B}^2 \Delta)\vec \varphi
\\
+ g \vec \varphi \cdot \psi^\dagger \gamma_0 \vec s \psi
+ \lambda \bigl( \vec \varphi \cdot \vec \varphi \bigr)^2 
+ \mathcal O(\varphi^6,\psi^4,\psi^2\varphi^4)\,,
\end{multline}
where $\psi \equiv \psi(\vec x, \tau) = k_{\mathrm B} T  \sum_{n} \int^\Lambda_0 \frac{\rmd^2 \vec k}{(2\pi a_\mathrm{M})^2} \rme^{\rmi \omega_n \tau + i \mathbf k \cdot \vec x}\psi_{\mathbf k}(\omega_n)$, with fermionic Matsubara frequencies $\hbar\omega_n = k_\mathrm{B} T (2n+1) \pi$ at temperature $T$ and an ultraviolet momentum cutoff $\Lambda \simeq 1/a_\mathrm{M}$.
The velocities $v_\mathrm{F}$ and $v_\mathrm{B}$ characterize the propagation speeds of the low-energy fermionic and magnetic excitations, respectively.
The tuning parameter $r$ varies linearly with $\theta-\theta_\mathrm{c}$ or $p_\mathrm{c} - p$ in the vicinity of the transition, while $g$ and $\lambda$ denote the Yukawa coupling and boson self-interaction, respectively.
The higher-order interaction terms indicated in Eq.~\eqref{eq:field-theory} are irrelevant in the renormalization-group sense and therefore do not influence the universal critical behavior.
The $\gamma_\mu$ are $4 \times 4$ Dirac matrices acting on minivalley and sublattice degrees of freedom and satisfy the Clifford algebra $\{\gamma_\mu,\gamma_\nu\}=2\delta_{\mu\nu}$.
In the reference frame defined by $\mathbf k_x = \mathbf k \cdot \boldsymbol{\kappa}/\kappa$ and $k_y = (\boldsymbol{\kappa} \times \mathbf k) \times \boldsymbol{\kappa}/\kappa^2$, a convenient representation of the Dirac matrices is $(\gamma_0,\gamma_1,\gamma_2) = (\tau_0 \sigma_z, \tau_z \sigma_y, \tau_0 \sigma_x)$~\cite{herbut06}. Here, $(\tau_x,\tau_y,\tau_z)$ denote Pauli matrices in minivalley space with $\tau_0$ the identity, while $(\sigma_x,\sigma_y,\sigma_z)$ act in sublattice space.
$\vec s=(s_x,s_y,s_z)$ denotes the Pauli matrices in spin space. 
For generic values of $v_\mathrm{F} \neq v_\mathrm{B}$, the Lagrangian lacks Lorentz symmetry. At the phase transition point, however, we show below that $v_\mathrm{F}$ and $v_\mathrm{B}$ flow to the same value at low energies, giving rise to emergent Lorentz invariance at criticality.

For $r>0$, the mean-field solution to the theory defined in Eq.~\eqref{eq:continuum-model} describes an SU(2)-symmetric phase with a vanishing order parameter, $\langle \vec\varphi \rangle = 0$, and an electronic spectrum featuring two gapless Dirac cones per spin.
For $r<0$, the order parameter acquires a finite expectation value  at zero temperature, $\langle \vec\varphi \rangle \neq 0$, which gaps out the Dirac cones and yields a phase with long-range antiferromagnetic order.

To describe the physics near the phase transition beyond mean-field theory, we employ a renormalization group analysis based on an $\epsilon$ expansion about the upper critical spatial dimension $d=3$~\cite{rosenstein93}.
As a starting point of this analysis, we estimate the values of the theory parameters at the moir\'e lattice scale $\Lambda \simeq 1/a_\mathrm{M}$.
From the spectrum of $\mathcal H_\text{kin}$ in Eq.~\eqref{eq:continuum-model}, we extract a Fermi velocity $v_\mathrm{F} \simeq 2.7 \times 10^4\,\mathrm{m/s}$ for $\theta \simeq 2.7^\circ$, in agreement to within 10\% with the value inferred from magnetotransport measurements in twisted double bilayer WSe$_2$ for $\theta \simeq 3^\circ$~\cite{ma24}.
To estimate $v_\mathrm{B}$, we consider an effective Hubbard model on the emergent honeycomb lattice in the strong-coupling limit, which yields an antiferromagnetic Heisenberg exchange coupling $J \simeq 4 t_1^2/U_0$~\cite{yang12}, where $t_1$ is the effective nearest-neighbor hopping amplitude and $U_0$ is the effective on-site interaction.
The magnetic excitation spectrum contains a gapless magnon mode with linear dispersion $\omega_{\mathbf k} = v_\mathrm{B} |\mathbf k| + \mathcal O(k^2)$ near the Brillouin-zone center. Within linear spin-wave theory, the corresponding magnon velocity is $v_\mathrm{B} \simeq \sqrt{3/2} a_\text{M} J S/\hbar$~\cite{auerbachbook}, where $a_\text{M} \simeq a_0/\theta$ is the moir\'e lattice constant and $S = 1/2$ is the spin length.
Within the effective Hubbard-model description, the Fermi velocity in the weak-coupling limit is $v_\mathrm{F} = \sqrt{3} t_1 a_\mathrm{M}/(2\hbar)$, which allows for a rough order-of-magnitude estimate of the spin-wave velocity relative to the Fermi velocity as
$v_\mathrm{B} 
\simeq 4\sqrt{2} (t_1/U_0) S v_\mathrm{F}
\simeq 2 \times 10^4\,\mathrm{m/s}$.
In the last step, we have used the critical ratio $(U_0/t_1)_\mathrm{c} \simeq 3.8$ for the semimetal-to-insulator transition of the honeycomb-lattice Hubbard model~\cite{assaad13}, together with the numerical value of $v_\mathrm{F}$ extracted from the spectrum of $\mathcal H_\text{kin}$.
The ratio $g^2/(2r)$ sets the effective electron-electron interaction scale responsible for the antiferromagnetic order. In Appendix~\ref{app:energy-scales}, we show that for $\theta = 2.7^\circ$ and $\epsilon_\text{eff} = 110$ this scale can be estimated as $g^2/(2r) \simeq U_{\mathrm S} \simeq 4.6\,\mathrm{meV}$.
The bosonic self-coupling $\lambda$ can be taken to be small at the moir\'e lattice scale; however, it is generated under renormalization group flow and must therefore be included to correctly capture the low-energy physics.

Integrating out modes in the momentum shell $\mathbf k \in [\Lambda/b,\Lambda]$ in $d = 3-\varepsilon$ spatial dimensions, together with all Matsubara frequencies $\omega \in (-\infty,\infty)$ at zero temperature, induces a renormalization group flow of the parameters in Eq.~\eqref{eq:continuum-model} along the critical surface $r=r_\star$, given by
\begin{align}
\frac{\rmd (v_\mathrm{F}/v_\mathrm{B})}{\rmd \ln b} & = (z-1-\eta_\psi) v_\mathrm{F}/v_\mathrm{B} + \frac{3}{8} f_p(v_\mathrm{F}/v_\mathrm{B}) g^2\,,
\displaybreak[0] \\
\frac{\rmd g^2}{\rmd \ln b} & = (\varepsilon + z - 1 - \eta_1 - 2 \eta_\psi) g^2 + \frac{1}{2} f_g(v_\mathrm{F}/v_\mathrm{B}) g^4\,,
\displaybreak[0] \\
\frac{\rmd \lambda}{\rmd \ln b} & = (\varepsilon + 3z - 3 - 2\eta_\phi)\lambda - 11 \lambda^2 + \frac{1}{2} f_\lambda(v_\mathrm{F}/v_\mathrm{B}) g^2\,,
\end{align}
where the dynamical critical exponent $z$ and the anomalous dimensions are defined as
\begin{align} \label{eq:anomalous-dimensions-1}
z & = 1 + \frac{\eta_\phi - \eta_1}{2}\,,
& 
\eta_\psi & = \frac{3}{8} f_\omega(v_\mathrm{F}/v_\mathrm{B}) g^2\,,
\\
\eta_\phi & = f_{\omega^2}(v_\mathrm{F}/v_\mathrm{B}) g^2\,,
&
\eta_1 & = f_{p^2}(v_\mathrm{F}/v_\mathrm{B}) g^2\,.
\end{align}
Away from the critical surface, the tuning parameter $r$ evolves as
\begin{align} \label{eq:flow-r}
\frac{\rmd r}{\rmd \ln b} & = (2 - \eta_1) r + 10 \frac{\lambda}{\sqrt{1+r}} - 4 f_r(v_\mathrm{F}/v_\mathrm{B}) g^2\,.
\end{align}
In the above equations, we have introduced dimensionless parameters according to 
$g^2 \Lambda^{1-z+\eta_1+2\eta_\psi-\varepsilon}/(2\pi^2) \mapsto g^2$,
$\lambda \Lambda^{3-3z+2\eta_\phi - \varepsilon}/(2\pi^2) \mapsto \lambda$,
and
$r \Lambda^{\eta_1 - 2} \mapsto r$.
The dimensionless cutoff functions $f_i$, with $i\in \{p,p^2,\omega,\omega^2,r,\lambda,g\}$, arise from the integration over Matsubara frequencies and are normalized such that $f_i(1)=1$. For the sharp momentum cutoff used here, they take the values $f_p(x) = 4(2+x)/[3(1+x)^2]$, $f_{p^2}(x) = f_r(x) = 1/x$, $f_\omega(x) = 4/(1+x)^2$, $f_{\omega^2}(x) = f_\lambda(x) = 1/x^3$, and $f_g = 2/[x(1+x)]$.

\begin{figure}[tb!]
\includegraphics[width=\linewidth]{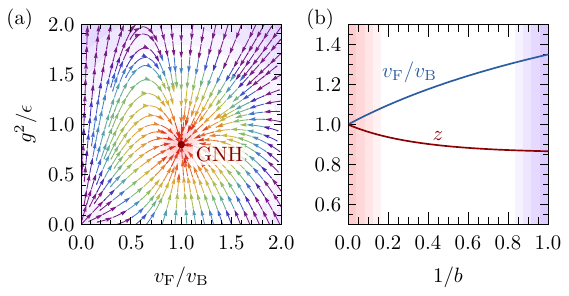}
\caption{%
(a)~Renormalization group flow in the plane spanned by the velocity ratio $v_\mathrm{F}/v_\mathrm{B}$ and the Yukawa coupling $g^2$, with the tuning parameter $r$ set to criticality, $r=r_\star$. Arrows indicate the flow toward the infrared. The stable fixed point at $v_\mathrm{F}/v_\mathrm{B} = 1$ and $g^2 = 4\varepsilon/5$ belongs to the Gross-Neveu-Heisenberg (GNH) universality class.
(b)~Velocity ratio $v_\mathrm{F}/v_\mathrm{B}$ (blue) and effective dynamical critical exponent $z$ (red) as functions of the renormalization group scale $1/b$ for $v_\mathrm{F}/v_\mathrm{B} = 1.35$ at the ultraviolet scale ($1/b = 1$). In the infrared limit ($1/b \to 0$), $v_\mathrm{F}/v_\mathrm{B}$ and $z$ both flow to unity, reflecting the emergence of Lorentz invariance at criticality.
}
\label{fig:7}
\end{figure}

The flow equations exhibit a unique critical fixed point at $g^2_\star = \frac{4}{5} \varepsilon + \mathcal O(\varepsilon^2)$ and $\lambda_\star = \frac{8}{55}\varepsilon + \mathcal O(\varepsilon^2)$, illustrated in Fig.~\ref{fig:7}(a).
Importantly, the fixed point is characterized by a single velocity,
\begin{align}
\frac{v_\mathrm{F}}{v_\mathrm{B}} & = 1\,,
\end{align}
so that both fermionic and magnetic excitations propagate at the same speed at low energies.
As a result, the system undergoes a continuous Dirac-semimetal-to-antiferromagnetic-insulator transition exhibiting emergent Lorentz invariance~\cite{ray21}.
This is illustrated in Fig.~\ref{fig:7}(b), which shows the evolution of the velocity ratio $v_\mathrm{F}/v_\mathrm{B}$ (blue) and the effective dynamical exponent $z$ from Eq.~\eqref{eq:anomalous-dimensions-1} (red) as functions of the renormalization group scale $1/b$, with $1/b = 1$ and $1/b \to 0$ corresponding to the ultraviolet and infrared limits, respectively.
Here, we initialize the flow with a velocity ratio consistent with the moir\'e-scale estimate $v_\mathrm{F}/v_\mathrm{B} = 1.35$, and choose the bare couplings $g^2=g_\star^2$ and $\lambda=0$ at the ultraviolet scale. Other choices of starting values produce qualitatively similar flows and the same infrared limit, provided that $r$ is tuned to criticality.
Using Eqs.~\eqref{eq:anomalous-dimensions-1}--\eqref{eq:flow-r} together with the fixed-point values, we obtain the critical exponents that characterize the universal behavior near the quantum critical point as
\begin{align}
z & = 1\,,
&
\nu & = \frac12 + \frac{21\varepsilon}{55} + \mathcal O(\varepsilon^2)\,,
\displaybreak[0] \\
\eta_\psi & = \frac{3\varepsilon}{10} + \mathcal O(\varepsilon^2)\,,
&
\eta_\phi & = \eta_1 = \frac{4\varepsilon}{5} + \mathcal O(\varepsilon^2)\,.
\end{align}
These values agree precisely with the established results for Gross-Neveu-Heisenberg universality to leading order in the $\varepsilon$ expansion about the upper critical dimension~\cite{herbut09, janssen14}.
This explicitly confirms that the semimetal-to-insulator transition in twisted double bilayer WSe$_2$ belongs to the Gross-Neveu-Heisenberg universality class.

To go beyond the leading-order $\varepsilon$ expansion, we can therefore rely on established results for Gross-Neveu-Heisenberg criticality to predict the quantum critical behavior.
The critical behavior of the Gross-Neveu-Heisenberg model has been studied extensively using a variety of analytical approaches, including the $\varepsilon$ expansion around the upper~\cite{herbut09, zerf17} and lower~\cite{ladovrechis23} critical dimensions, the large-$N$ expansion~\cite{gracey18}, and the functional renormalization group~\cite{janssen14,knorr18,tolosasimeon25}. Complementary numerical investigations have explored the transition using different lattice realizations~\cite{%
assaad13, toldin15, liu19, liu21-gnh,
otsuka16, otsuka20,
xu21,
buividovich18, buividovich19, ostmeyer20, ostmeyer21,
lang19}.

These results enable predictions of the universal power laws governing the thermodynamic, spectroscopic, and transport properties near the quantum critical point.
In the antiferromagnetic phase, the staggered magnetization scales with twist angle or pressure as
$\langle \vec \varphi \rangle \propto (\theta_\mathrm{c} - \theta)^\beta$ or $\langle \vec \varphi \rangle \propto (p - p_\mathrm{c})^\beta$,
with nonuniversal prefactors but a universal order-parameter exponent $\beta \approx 1.21$.
The electronic gap exhibits the scaling
$\Delta \propto (\theta_\mathrm{c} - \theta)^{\nu z}$ or $\Delta \propto (p - p_\mathrm{c})^{\nu z}$,
with dynamical exponent $z = 1$ and correlation-length exponent $\nu \approx 1.20$.
At the quantum critical point $\theta = \theta_\mathrm{c}$ or $p = p_\mathrm{c}$, the fermionic (magnetic) excitation spectrum will be characterized by a continuum of low-energy excitations near the $\boldsymbol \kappa$ ($\boldsymbol \gamma$) point of the moir\'e Brillouin zone, with lower bounds
\begin{align}
\varepsilon_{\mathbf k, \text F} & \geq v_\mathrm{F} |\mathbf k - \boldsymbol{\kappa}| + \mathcal O((\mathbf k - \boldsymbol{\kappa})^2)\,,\\
\varepsilon_{\mathbf k, \text B} & \geq v_\mathrm{B} |\mathbf k| +  \mathcal O(\mathbf k^2)\,.
\end{align}
Importantly, in the low-energy limit the velocities $v_\mathrm{F}$ and $v_\mathrm{B}$ flow to the same value, reflecting the emergent Lorentz invariance at criticality.
Furthermore, the dynamical spin structure factor at the quantum critical point scales as
$\mathcal S(\mathbf k, \omega) \propto 1/{(\omega^2 - \mathbf k^2)^{(2-\eta_\varphi)/2}}$,
with order-parameter anomalous dimension $\eta_\varphi \approx 1.01$.
The quoted exponents are estimated from interpolations between expansions near the lower and upper critical dimensions~\cite{ladovrechis23}, using standard hyperscaling relations~\cite{herbutbook}.

\subsection{Strained samples}

\begin{figure}[tb!]
\includegraphics[trim={0 0.4cm 0 0},width=\linewidth]{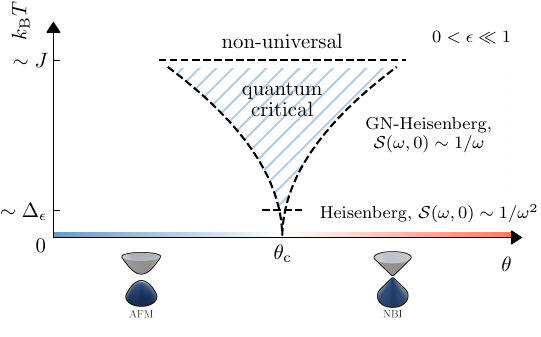}
\caption{
Schematic phase diagram of weakly-strained twisted double bilayer TMDs as a function of temperature~$T$ and twist angle~$\theta$.
For temperatures $k_\mathrm{B} T$ below the strain-induced gap $\Delta_\epsilon$, the critical behavior is governed by the (2+1)D Heisenberg universality class.
At intermediate temperatures above $\Delta_\epsilon$ and below the microscopic interaction scale $J$, observables instead follow Gross-Neveu-Heisenberg criticality, characterized by significantly larger values of $\eta_\varphi$, $\nu$, and $\beta$.
This crossover may be revealed, for example, in the dynamic spin structure factor at zero momentum, which changes from $\mathcal S(\omega,0)\sim 1/\omega^2$ scaling for $k_\mathrm{B} T \ll \Delta_\epsilon$ to $\mathcal S(\omega,0)\sim 1/\omega$ scaling for $\Delta_\epsilon \ll k_\mathrm{B} T \ll J$.
For twisted double bilayer WSe$_2$ with strain magnitude $\epsilon = 0.5\%$, we estimate $J/k_\mathrm{B} \sim 200\,\text{K}$ and $\Delta_\epsilon/k_\mathrm{B} \sim 10\,\text{K}$.
}
\label{fig:8}
\end{figure}

Finite heterostrain breaks the $C_{2y}$ symmetry of the microscopic model. In the field-theory description, this corresponds to an explicit Dirac mass term $m \bar\psi \psi$ in the continuum Lagrangian, which gaps out the spectrum already in the noninteracting limit.
Consequently, the Dirac fermions can be integrated out without singularities, leaving a purely bosonic low-energy theory in the form of the standard O(3) model, falling into the (2+1)D Heisenberg universality class~\cite{herbutbook},
\begin{align}
S_\mathrm{H} = \int \rmd \tau \rmd^2 \vec x \,\biggl[
\frac{1}{2} ( \partial_\mu \vec \varphi )^2
+ \frac{r}{2} \vec \varphi^2 + \lambda \bigl( \vec \varphi^2 \bigr)^2 \biggr]\,.
\end{align}
We emphasize that the O(3) model above is valid only at energies below the strain-induced single-particle gap $\Delta_\epsilon$.
For weakly strained samples, $\Delta_\epsilon$ is small compared to the microscopic interaction scale $J$, which sets the order of the interaction-induced band gap $\Delta$ deep in the ordered phase. 
In the strong-coupling regime, this scale can be estimated as $J \sim 4 \overline{t}^2 /U_\text{S}$, where $\overline{t}$ is the $\mathbf k$-averaged bandwidth and $U_\text{S}$ is the effective interaction strength responsible for the antiferromagnetic order, see Appendix~\ref{app:energy-scales}.
For twist angles near the critical value, $\theta \simeq \theta_\mathrm{c} = 2.7^\circ$, we estimate $\overline{t} \simeq 4.8\,\text{meV}$ and $U_\mathrm{S} \simeq 4.6\,\text{meV}$ for $\epsilon_\text{eff} = 110$, giving $J \simeq 20\,\text{meV}$ for the microscopic energy scale.
At energies above~$\Delta_\epsilon$ but below~$J$, such systems remain governed by Gross-Neveu-Heisenberg universality, described by Eq.~\eqref{eq:field-theory}.
This leads to a crossover within the quantum critical regime at finite temperatures $T$ above the quantum critical point at $\theta = \theta_\mathrm{c}$ or $p = p_\mathrm{c}$, as schematically shown in Fig.~\ref{fig:8}.
%
%
For $k_\mathrm{B} T \ll \Delta_\epsilon$, the critical behavior is governed by the (2+1)D Heisenberg universality class, with anomalous dimension $\eta_\varphi \approx 0.038$, correlation-length exponent $\nu \approx 0.71$, and order-parameter exponent $\beta \approx 0.37$~\cite{campostrini02, kompaniets17, cabrera25}.
For $\Delta_\epsilon \ll k_\mathrm{B} T \ll J$, observables instead follow (2+1)D Gross-Neveu-Heisenberg criticality, with exponents $\eta_\varphi$, $\nu$, and $\beta$ given in the previous subsection, which are significantly larger than their Heisenberg-class counterparts.
This crossover may be revealed, for example, in the dynamic spin structure factor at zero momentum, which changes from $\mathcal S(\omega,0)\sim 1/\omega^2$ scaling for $k_\mathrm{B} T \ll \Delta_\epsilon$ to $\mathcal S(\omega,0)\sim 1/\omega$ scaling for $\Delta_\epsilon \ll k_\mathrm{B} T \ll J$.
At even higher temperatures, $k_\mathrm{B} T \gg J$, the behavior becomes nonuniversal and depends on microscopic details.

\section{Conclusions}
\label{sec:conclusions}

We have investigated the ground-state phase diagram of $\boldsymbol{\Gamma}$-valley double bilayer TMDs at hole filling $\nu = 2$ per moiré unit cell as a function of twist angle and applied pressure.
At large twist angles, the system realizes a fully symmetric Dirac semimetal with spin-degenerate Dirac cones located at the moiré-Brillouin-zone corners $\boldsymbol{\kappa}$ and $\boldsymbol{\kappa}’$.
Reducing the twist angle drives a continuous transition into a Néel-type antiferromagnetic insulator on an emergent honeycomb lattice formed by the AB and BA stacking regions of the inner TMD layers.
Unlike conventional antiferromagnets, where spin-density modulations occur on the microscopic lattice scale, the ordering wavelength of this antiferromagnetic state is set by the moiré scale.
At yet smaller twist angles, a level crossing occurs from the antiferromagnetic insulator to a ferromagnetic insulator with spin-split bands, stabilized by antichiral terms in the Hamiltonian.

We have argued that the continuous transition from the Dirac semimetal to the antiferromagnetic insulator exhibits emergent relativistic symmetry and belongs to the (2+1)D Gross-Neveu-Heisenberg universality class.
For samples with twist angles just above the critical value at ambient conditions, uniaxial or hydrostatic pressure provides a highly sensitive and experimentally accessible tuning knob to drive the semimetal-to-insulator transition.
With finite heterostrain, the Dirac cones are gapped already in the noninteracting limit. Consequently, the critical behavior at the lowest temperatures follows conventional (2+1)D Heisenberg universality, crossing over to (2+1)D Gross-Neveu-Heisenberg behavior above the strain-induced gap, provided this gap is small compared to the microscopic energy scale.

These theoretical predictions can be directly tested in hole-doped twisted double bilayer TMDs with a valence band edge at the $\boldsymbol{\Gamma}$ point.
In particular, our results suggest that the insulating state observed experimentally in twisted double bilayer WSe$_2$~\cite{ma24} is a Néel-type antiferromagnet with a spin-density wavelength set by the moiré scale.
This substantial enhancement of the wavelength compared to conventional Néel antiferromagnets may enable direct imaging of the spin-density modulation using advanced nanoscale magnetometry techniques, including spin-resolved scanning tunneling microscopy~\cite{bergmann14}, nitrogen-vacancy-center magnetometry~\cite{hong13}, and potentially nanoSQUID devices~\cite{martinez17, rog25} or advanced magnetic force microscopy~\cite{schmidt24}.
Pressure experiments on samples with twist angles just above the ambient-condition critical angle could further test our predictions for the power-law scaling of the order parameter, $m_\text{AFM} \propto (p - p_\mathrm{c})^\beta$, with an unusually large exponent $\beta \approx 1.21$ compared to the conventional (2+1)D Heisenberg universality class.
Transport experiments could likewise probe the predicted gap scaling, $\Delta \propto (p - p_\mathrm{c})^{\nu z}$, with $\nu \approx 1.20$ and $z = 1$.
Ultrafast optical techniques~\cite{wu24, mitra25} may enable tests of the predicted crossover from Gross-Neveu-Heisenberg to Heisenberg universality, reflected in the frequency scaling of the dynamical spin structure at low temperatures in weakly strained samples.

From a broader perspective, two additional remarks are in order.
First, quantum critical points involving gapless fermionic excitations have been discussed previously in a wide range of strongly-correlated systems, including cuprate and iron-based superconductors, heavy-fermion compounds, and other moir\'e materials.
In these settings, however, the absence of a sufficiently high symmetry has precluded a controlled theoretical description of the universal physics governing the transition.
By contrast, the emergent Lorentz symmetry that we predict at the semimetal-to-insulator transition in twisted double bilayer WSe$_2$ enables a field-theoretical treatment that is fully accessible within a standard $\varepsilon$ expansion about the upper critical spatial dimension $d=3$.
The quantum critical point in twisted double bilayer WSe$_2$ may therefore constitute the first experimentally realizable example of a fermionic quantum universality class that is under complete theoretical control.
Second, a variant of the Gross-Neveu-Heisenberg model identified here, incorporating randomized couplings, has recently been shown to support a superconducting dome emerging from strong critical fluctuations as a secondary instability above the quantum critical point~\cite{stangier25a, stangier25b}.
While we do not expect such a secondary instability to arise in pristine twisted double bilayer WSe$_2$, an important direction for future work is to identify the conditions under which randomized Gross-Neveu-Heisenberg criticality may be realized, potentially offering new insights into the emergence of strong-coupling superconductivity above quantum critical points.

\paragraph*{Note added.}
After this manuscript was submitted, a simultaneous study appeared on arXiv, which employs a Hubbard-type model on the emergent honeycomb lattice as an effective description of twisted double bilayer WSe$_2$~\cite{hawashin25b}.
 
\begin{acknowledgments}
%
We thank
Bilal Hawashin,
Julian Kleeschulte,
David Kurz,
and
Michael Scherer
for valuable discussions.
This work has been supported by the Deutsche Forschungsgemeinschaft (DFG) through 
Project No.\ 247310070 (SFB 1143, A07), 
Project No.\ 390858490 (W\"urzburg-Dresden Cluster of Excellence \textit{ctd.qmat}, EXC 2147), and 
Project No.\ 411750675 (Emmy Noether program, JA2306/4-1).

\end{acknowledgments}

\section*{Data availability}

The data that support the findings of this article are openly available~\cite{data-availability}. 

\appendix
\setcounter{equation}{0}
\renewcommand\theequation{A\arabic{equation}}

\appsection{Interacting continuum model}
\label{app:model}

This appendix provides further details of the interacting continuum model introduced in Sec.~\ref{sec:model}.
The noninteracting part of the Hamiltonian, repeated here for convenience, has the form
\begin{align} \label{eq:continuum-model-app}
\mathcal{H}_{\mathrm{kin}} &= \sum_{\mathbf q} f^\dagger_{\mathbf q} \left(h_{\boldsymbol{\Gamma}}(\mathbf q) + h_0 \right) f_{\mathbf q} +
\sum_{\mathbf q} \sum_{j = 1}^6 f^\dagger_{\mathbf q + \mathbf G_j} h_1(\mathbf G_j) f_{\mathbf q},
\end{align}
with $4\times 4$ matrices $h_0$ and $h_1(\mathbf G_j)$ acting in layer space. They are given by
\begin{align}
h_0 &=
\begin{pmatrix}
V_1 & V_{12} & 0 & 0 \\
V_{12} & V_{2}^{(0)} & V_{23}^{(0)} & 0 \\
0 & V_{23}^{(0)} & V_{2}^{(0)}  & V_{12} \\
0 & 0 & V_{12} & V_{1} \\
\end{pmatrix},
\end{align}
and
\begin{align}
h_1(\mathbf G) &=
\begin{pmatrix}
0 & 0 & 0 & 0 \\
0 & V_{2}^{(1)} \rme^{\rmi \varphi \operatorname{sgn}(\mathbf G)} & V_{23}^{(1)} & 0 \\
0 & V_{23}^{(1)} & V_{2}^{(1)} \rme^{-\rmi \varphi \operatorname{sgn}(\mathbf G)}  & 0 \\
0 & 0 & 0 & 0 \\
\end{pmatrix}.
\end{align}
Here, we have defined $\operatorname{sgn}(\boldsymbol G_{1, 3, 5}) = +1$ and $\operatorname{sgn}(\boldsymbol G_{2, 4, 6}) = -1$ for the reciprocal lattice vectors $\boldsymbol G_j$ indicated in Fig.~\ref{fig:1}.
The intralayer tunneling amplitudes $(V_1, V_2^{(0)}, V_2^{(1)}) = (200, -159, -8)\,\mathrm{meV}$, phase shift $\varphi = -0.17$, and interlayer tunneling amplitudes $(V_{12}, V_{23}^{(0)}, V_{23}^{(1)}) = (184, 356, -9)\,\mathrm{meV}$, realistic for twisted double bilayer WSe$_2$, are adopted from Ref.~\cite{pan23}.
In order to diagonalize the Hamiltonian, we expand the eigenfunctions in a basis of plane waves as
\begin{align}
\label{eq:bloch-states}
\ket{\psi_{\mathbf k, n}(\mathbf r)} = \frac{1}{\sqrt{(2\pi)^2}} \sum_{\substack{\mathbf G \\ \vert \mathbf G \vert \leq 4 \vert \mathbf G_{1, 2} \vert}} \sum_{\ell} u_{\mathbf k, n; \mathbf G, \ell} \rme^{\rmi(\mathbf k + \mathbf G) \cdot \mathbf r} \ket{\ell},
\end{align}
where the sum over reciprocal lattice vectors $\mathbf G$ is truncated to a circle of radius 4$\vert \mathbf G_{1, 2} \vert$ for the numerical diagonalization, yielding 244 energy bands per spin species.

Our numerics is performed in the eigenbasis of the continuum Hamiltonian, with fermionic operators
\begin{align}
c^\dagger_{\mathbf k, n, s} = \sum_{\mathbf G, \ell} u_{\mathbf k, n; \mathbf G, \ell} f^\dagger_{\mathbf k + \mathbf G, \ell, s}.
\end{align}
In the continuum description, the fermionic operators in momentum space are related to those in real space via the Fourier transformation
\begin{align}
f^\dagger_{\mathbf R, \ell, s} = \frac{1}{\sqrt A} \sum_{\mathbf q} \rme^{\rmi \mathbf q \cdot \mathbf R} f^\dagger_{\mathbf q, \ell, s},
\end{align}
and hence real space expectation values are given by
\begin{align}
\langle f^\dagger_{\mathbf R, \ell, s} f_{\mathbf R, \ell^\prime, s^\prime} \rangle &= \frac{1}{A} \sum_{\mathbf k, \mathbf G, \mathbf G^\prime} \sum_{n, n^\prime} \rme^{\rmi(\mathbf G - \mathbf G^\prime) \cdot \mathbf R} u_{\mathbf k, n; \mathbf G, \ell}^* u_{\mathbf k, n^\prime; \mathbf G^\prime, \ell^\prime} \nonumber \\
&\quad\times \langle c^\dagger_{\mathbf k, n, s} c_{\mathbf k, n^\prime, s^\prime} \rangle.
\end{align}

When taking into account Coulomb interactions, we truncate the band basis to the $N_{\text{bands}}$ topmost bands per spin species for computational purposes~\cite{bultinck20, liu21}.
The remaining remote energy bands are assumed to be fully occupied, with their electrons not contributing substantially to the low-energy physics due to their distance to the Fermi level.
Unless stated otherwise, we use a value of $N_\text{bands} = 6$ in our numerics.
The Coulomb potential $V_{\mathbf q}$ is truncated at a radius of 2$\vert \mathbf G_{1, 2} \vert$. 

An additional subtlety arises in the implementation of the Coulomb interaction:
The parameters of the continuum Hamiltonian in Eq.\eqref{eq:continuum-model-app} are obtained by fitting a tight-binding model~\cite{pan23} to ab initio calculations~\cite{fang15}.
Consequently, the single-particle dispersion of the continuum model already incorporates interaction effects present in the fully symmetric ground state, which must be removed using a subtraction scheme~\cite{bultinck20, faulstich23, kwan22}.
The ab initio results are computed at charge neutrality, where the fully symmetric density matrix is  $P_{\text{ref}} = \mathbb{1}_{N_{\text{bands}}}$.
To obtain the bare dispersion, we subtract interaction effects due to this density matrix from the continuum-model dispersion.
Denoting the first-quantized kinetic Hamiltonian in the band basis by $h_0(\mathbf k)$, this amounts to replacing $h_0(\mathbf k) \to h(\mathbf k) = h_0(\mathbf k) - h_{\mathrm{HF}}[P_{\text{ref}}](\mathbf k)$. This subtraction scheme is applied in the interacting Hamiltonian of Eq.~\eqref{eq:full-model}.
Note that remote-band contributions to the interacting Hamiltonian due to $P$ and $P_{\text{ref}}$ cancel for our choice of subtraction scheme, meaning it suffices to consider the effective Hamiltonian in the space of the active bands.

\setcounter{equation}{0}
\renewcommand\theequation{B\arabic{equation}}
\appsection{Strong-coupling analysis}
\label{app:strong-coupling}

This appendix provides further details of the strong-coupling analytics outlined in Sec.~\ref{sec:strong-coupling}.

\subsection{Overlap matrices}

Our calculations are based on a decomposition of the overlap matrices $\Lambda(\mathbf k, \mathbf q) = \Lambda_{\mathrm{S}}(\mathbf k, \mathbf q) + \Lambda_{\mathrm{A}}(\mathbf k, \mathbf q)$ into simple symmetric and antisymmetric contributions $\Lambda_{\mathrm{S / A}}(\mathbf k, \mathbf q)$ with respect to the chiral symmetry operator $\sigma_z$~\cite{bultinck20}:
The combined $C_{2z} \mathcal T$ symmetry implies that the overlap matrices in the sublattice basis may contain terms involving $\sigma_{0}=\mathbb 1_2$ and $\sigma_x$ with real coefficients and terms involving $\sigma_{y}$ and $\sigma_z$ with imaginary coefficients.
Only the $\sigma_{0}$ and $\sigma_z$ terms are symmetric under chiral transformations, such that the chiral component of the overlap matrices has the form
\begin{align}
\Lambda_{\mathrm S }(\mathbf k, \mathbf q) = (a_0(\mathbf k, \mathbf q) \sigma_0 + a_x(\mathbf k, \mathbf q) \rmi \sigma_z) s_0,
\end{align}
where $a_{0}(\mathbf k, \mathbf q)$ and $a_{x}(\mathbf k, \mathbf q)$ are real-valued functions, and the trivial spin dependence $\propto s_0$ is enforced by SU(2) symmetry.
Suppressing the trivial spin structure, this is equivalent to the expression
\begin{align}
\Lambda_{\mathrm S }(\mathbf k, \mathbf q) = F_{\mathrm S}(\mathbf k, \mathbf q) \rme^{\rmi \Phi_{\mathrm S}(\mathbf k, \mathbf q) \sigma_z}
\end{align}
given in the main text.
The remaining $\sigma_{x}$ and $\sigma_y$ terms make up the antichiral contribution, analogously yielding
\begin{align}
\Lambda_{\mathrm A}(\mathbf k, \mathbf q) & = F_{\mathrm A}(\mathbf k, \mathbf q) \sigma_x \rme^{\rmi \Phi_{\mathrm A}(\mathbf k, \mathbf q) \sigma_z}.
\end{align}
Time-reversal and $C_{2z}$ rotation symmetry both separately require that
\begin{align} \label{eq:trs_f}
F_{\mathrm S / \mathrm A}(-\mathbf k, -\mathbf q) = F_{\mathrm S / \mathrm A}(\mathbf k, \mathbf q)
\end{align}
and, up to multiples of $2\pi$,
\begin{align} \label{eq:trs_phi}
\Phi_{\mathrm S / \mathrm A}(-\mathbf k, -\mathbf q) = -\Phi_{\mathrm S / \mathrm A}(\mathbf k, \mathbf q).
\end{align}
The definition of the overlap matrices additionally implies the identities
\begin{align} \label{eq:lambda-dagger}
\Lambda(\mathbf k, \mathbf q)^\dagger = \Lambda(\mathbf k + \mathbf q, -\mathbf q),
\end{align}
and, assuming a periodic gauge $u_{\mathbf k + \mathbf G^\prime, n; \mathbf G, \ell} = u_{\mathbf k, n; \mathbf G + \mathbf G^\prime, \ell}$ for the wavefunctions,
\begin{align} \label{eq:lambda-dagger-periodic}
\Lambda(\mathbf k, \mathbf G)^\dagger = \Lambda(\mathbf k, -\mathbf G).
\end{align}

\subsection{Strong-coupling Hamiltonian}

Following Ref.~\cite{bultinck20}, we now rewrite the interacting Hamiltonian in a form suitable for analytical arguments.
As discussed in Appendix~\ref{app:model}, the bare dispersion $h(\mathbf k)$ in Eq.~\eqref{eq:full-model} has the form $h(\mathbf k) =  h_0(\mathbf k) - h_{\mathrm{HF}}[P_{\text{ref}}](\mathbf k)$ with reference density matrix $P_{\text{ref}}$.
Rewriting $P_{\text{ref}} = \frac{1}{2}(\mathbb{1}_4 + Q_{\text{ref}})$, we may express the dispersion as
\begin{align}
h(\mathbf k) &= h_0(\mathbf k) - h_{\mathrm{HF}}[\tfrac12 Q_{\text{ref}}](\mathbf k) - h_{\mathrm{HF}}[\tfrac12 \mathbb{1}_4](\mathbf k) \\
&= \tilde h(\mathbf k) - h_{\mathrm{HF}}[\tfrac12 \mathbb{1}_4](\mathbf k).
\end{align}
The first term is now treated as renormalized dispersion, whereas the second term is absorbed into the interaction, yielding the expression
\begin{align}
\mathcal H &= \sum_{\mathbf k} c^\dagger_{\mathbf k} \tilde h(\mathbf k) c_{\mathbf k} - \frac{1}{2A} \sum_{\mathbf q} V_{\mathbf q} :\mathrel{{\rho_{\mathbf q} \rho_{-\mathbf q}}}: - \sum_\mathbf k c^\dagger_{\mathbf k} h_{\mathrm{HF}}[\tfrac12 \mathbb{1}](\mathbf k) c_{\mathbf k}
\end{align}
for the interacting Hamiltonian.
The normal-ordered density-density operator can be expressed as
\begin{align}
- \frac{1}{2A} \sum_{\mathbf q} V_{\mathbf q} :\mathrel{\rho_{\mathbf q} \rho_{-\mathbf q}}: \, = \frac{1}{2A} \sum_{\mathbf q} V_{\mathbf q} \rho_{\mathbf q} \rho_{-\mathbf q} + \sum_\mathbf k c^\dagger_{\mathbf k} h_{\mathrm F}[\tfrac12 \mathbb{1}](\mathbf k) c_{\mathbf k},
\end{align}
with
\begin{align}
\sum_\mathbf k c^\dagger_{\mathbf k} h_{\mathrm H}[\tfrac12 \mathbb{1}](\mathbf k) c_{\mathbf k} = \frac{1}{2A} \sum_{\mathbf q} V_{\mathbf q} \left( \bar{\rho}_{\mathbf q} \rho_{-\mathbf q} + \bar{\rho}_{-\mathbf q} \rho_{\mathbf q} \right).
\end{align}
Combining these expressions yields the strong-coupling form of the interacting Hamiltonian in Eq.~\eqref{eq:strong-coupling-model} of the main text, with the additive constant given by
\begin{align}
\text{const.} = -\frac{1}{2A} \sum_{\mathbf q} V_{\mathbf q} \bar{\rho}_{\mathbf q} \bar{\rho}_{-\mathbf q}.
\end{align}

As outlined in the main text, we begin the strong-coupling analysis by finding Slater determinant ground states of the dominant contribution $\mathcal H_{\mathrm S}$ to the Hamiltonian.
This is accomplished by noting that the charge density operator $ \delta \rho^{\mathrm S}_{\mathbf q}$ satisfies $\delta \rho^{\mathrm S}_{-\mathbf q} = (\delta \rho^{\mathrm S}_{\mathbf q})^\dagger$ due to the properties of the overlap matrices [Eqs.~\eqref{eq:lambda-dagger} and~\eqref{eq:lambda-dagger-periodic}], implying that $\mathcal H_{\mathrm S}$ is a positive semidefinite operator and any many-body state annihilated by $\mathcal H_{\mathrm S}$ must be a ground state in the chiral limit~\cite{bultinck20}.

\subsection{Energy scales}
\label{app:energy-scales}

To confirm the validity of our strong-coupling analysis, we numerically estimate the relevant energy scales. To this end, we explicitly evaluate the chiral part $\Lambda_{\mathrm S}(\mathbf k, \mathbf q)$ and the antichiral part $\Lambda_{\mathrm A}(\mathbf k, \mathbf q)$ of the overlap matrices, their magnitudes $F_{\mathrm S}(\mathbf k, \mathbf q)$ and $F_{\mathrm A}(\mathbf k, \mathbf q)$, and the corresponding interaction energy scales $U_S$ and $U_A$.
The latter are then compared with the average bandwidth $\bar t$, which represents the effective kinetic energy scale.

Note that the definition of the quantities $\Lambda_{\mathrm S / \mathrm A}(\mathbf k, \mathbf q)$ involves the sublattice operator $\sigma_z$, but there exists no explicit expression for the matrix elements of this operator in the band basis, because the sublattice degree of freedom is an emergent property.
In our numerics, we circumvent this problem by extracting the sublattice operator $\sigma_z$ from the density matrix $P_{\text{AFM}}$ of an antiferromagnetic state: In the absence of dispersion, converging the antiferromagnetic state yields a density matrix of the form $P_{\text{AFM}} = \frac{1}{2}(\mathbb{1} + \sigma_z \mathbf n \cdot \mathbf s)$, where we can fix $\mathbf n = \mathbf e_z$ by choosing a suitable initialization and then solve for $\sigma_z$.

We extract the magnitudes $F_{\mathrm S / \mathrm A}(\mathbf k, \mathbf q)$ of the overlap matrices from the basis-independent expression
\begin{align}
F_{\mathrm S / \mathrm A}(\mathbf k, \mathbf q)^2 = \frac{1}{4} \Tr \left[ \Lambda_{\mathrm S / \mathrm A}(\mathbf k, \mathbf q) \Lambda_{\mathrm S / \mathrm A}(\mathbf k, \mathbf q)^\dagger \right].
\end{align}
The corresponding energy scales $U_{\mathrm S / \mathrm A}$ of the interaction terms $\mathcal H_{\mathrm S / \mathrm A}$ may then be estimated by~\cite{bultinck20}
\begin{align}
U_{\mathrm S} &= \frac{1}{2AL^2} \sum_{\mathbf k, \mathbf q} V_{\mathbf q} F_{\mathrm S}(\mathbf k, \mathbf q)^2, \\
U_{\mathrm A} &= \frac{1}{2AL^2} \sum_{\mathbf k, \mathbf q} V_{\mathbf q} \left(2 F_{\mathrm S}(\mathbf k, \mathbf q) F_{\mathrm A}(\mathbf k, \mathbf q) + F_{\mathrm A}(\mathbf k, \mathbf q)^2 \right).
\end{align}
These quantities are compared to the $\mathbf k$-averaged bandwidth $\bar t$ of the renormalized dispersion $\tilde h(\mathbf k)$, given by
\begin{align}
\bar t 
= \frac{1}{2L^2} \sum_{\mathbf k} \left( \varepsilon_+(\mathbf k) - \varepsilon_-(\mathbf k) \right),
\end{align}
where $\varepsilon_\pm(\mathbf k)$ is the upper (lower) band in the spectrum of $\tilde h(\mathbf k)$.

\begin{figure}[tbp]
\includegraphics[width=\linewidth]{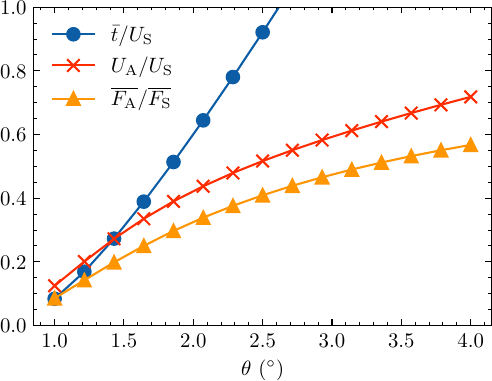}
\caption{%
Kinetic energy scale $\bar t$ (blue dots) and antichiral interaction energy scale $U_{\mathrm A}$ (red crosses), both normalized to the chiral interaction energy scale $U_{\mathrm S}$, together with the ratio of the average magnitudes $\overline{F_{\mathrm A}}/\overline{F_{\mathrm S}}$ of the antichiral and chiral components of the overlap matrices (orange triangles).
Calculations were performed using a fixed effective permittivity $\epsilon_\text{eff} = 110$, on an $18 \times 18$ momentum-space grid, keeping two bands per spin species. At small twist angles, the chiral interaction dominates over both the antichiral contribution and the dispersion, justifying a perturbative treatment.
Near the critical twist angle $\theta \simeq 2.7^\circ$, the kinetic energy scale $\overline t$ and the interaction energy scale $U_\mathrm{S}$ are comparable, as expected.}
\label{fig:9}
\end{figure}

Figure~\ref{fig:9} displays the $\mathbf k$-averaged bandwidth $\bar t$ and the antichiral interaction energy scale $U_{\mathrm A}$, both normalized to the chiral interaction energy scale $U_{\mathrm S}$, together with the ratio of the average magnitudes $\overline{F_{\mathrm A}}/\overline{F_{\mathrm S}}$ of the antichiral and chiral components of the overlap matrices. Here, the averages are defined as $\overline{F_{\mathrm S/A}} = \sum_{\mathbf k, \mathbf q} F_{\mathrm S/A}(\mathbf k, \mathbf q) / L^4$.
These results confirm that for small twist angles, both the dispersion and the antichiral interaction can indeed be treated as small perturbations to the chiral Hamiltonian $\mathcal H_\mathrm S$.

\subsection{Hartree-Fock treatment}

The ground states we consider in the strong-coupling limit are single Slater determinant states.
We may therefore evaluate their energies using Hartree-Fock theory.
For a given reduced density matrix $P$, this amounts to calculating the Hartree-Fock energy
\begin{align}
E_{\mathrm{HF}, \mathrm S}[P] &= \langle \mathcal H_{\mathrm S} \rangle \\
&=
\frac{1}{2A} \sum_{\mathbf q} V_{\mathbf q} \Bigl(
\langle \rho^{\mathrm S}_{\mathbf q} \rho^{\mathrm S}_{-\mathbf q} \rangle - \bar{\rho}^{\mathrm S}_{\mathbf q} \langle \rho^{\mathrm S}_{-\mathbf q} \rangle - \bar{\rho}^{\mathrm S}_{-\mathbf q} \langle \rho^{\mathrm S}_{\mathbf q} \rangle 
\nonumber \\ & \quad
+ \bar{\rho}^{\mathrm S}_{\mathbf q} \bar{\rho}^{\mathrm S}_{-\mathbf q} 
\Bigr)
\end{align}
where the expectation values are taken with respect to the many-body state $\ket{\Psi_P}$ parametrized by $P$.
This yields
\begin{align} \label{eq:hf-energy-sym}
E_{\mathrm{HF}, \mathrm S}[P] = \frac{1}{2} \sum_{\mathbf k} \Tr \left\lbrace P h_{\mathrm{HF}}^{\mathrm S}[P - \mathbb{1}_{4}](\mathbf k) \right\rbrace + \Delta E_{\mathrm S},
\end{align}
with the shorthand $h_{\mathrm{HF}}^{\mathrm S}[P](\mathbf k) = h_{\mathrm{H}}^{\mathrm S}[P](\mathbf k) + h_{\mathrm{F}}^{\mathrm S}[P](\mathbf k)$, and $h_{\mathrm{H}}^{\mathrm S}$, $h_{\mathrm{F}}^{\mathrm S}$ the Hartree and Fock contributions evaluated in the chiral limit $\Lambda(\mathbf k, \mathbf q) \to \Lambda_{\mathrm S}(\mathbf k, \mathbf q)$.
The additive constant is given by
\begin{align}
\Delta E_{\mathrm S} &= \frac{1}{2A} \sum_{\mathbf q} V_{\mathbf q} \bar{\rho}^{\mathrm S}_{\mathbf q} \bar{\rho}^{\mathrm S}_{-\mathbf q}
\\
&= \frac{2}{A} \sum_{\mathbf G } V_{\mathbf G} \left(\sum_{\mathbf k} F_{\mathrm S}(\mathbf k, \mathbf G) \cos \Phi_{\mathrm S}(\mathbf k, \mathbf G) \right)^2.
\end{align}

\subsection{Energy functional}

Due to the SU(2) symmetry of the Hamiltonian, we may simplify the computation of the Hartree-Fock energy by taking $\mathbf n \cdot \mathbf{s} \to s_z$ in the following for the otherwise arbitrary unit vector $\mathbf n$ in Eq.~\eqref{eq:candidate-states}.
To proceed with the energy evaluation, we first note that the Fock contribution to Eq.~\eqref{eq:hf-energy-sym},
\begin{align}
E_{\mathrm F, \mathrm S}[P] &= \frac{1}{2} \sum_{\mathbf k} \Tr \left\lbrace P h_{\mathrm F}^{\mathrm S}[P - \mathbb{1}_4] \right\rbrace,
\end{align}
vanishes for each of the three candidate ground states given in the main text:
Since both the overlap matrices $\Lambda_{\mathrm S}(\mathbf k, \mathbf q)$ and the density matrices $P$ under consideration are diagonal in the sublattice basis, they commute, yielding $h_{\mathrm F}^{\mathrm S}[P - \mathbb{1}_4](\mathbf k) \propto P - \mathbb{1}_4$.
The Fock contribution is thus proportional to $\Tr \left\lbrace P (P - \mathbb{1}_4) \right\rbrace = 0$, where we have used that $P^2 = P$.

The Hartree contribution evaluates to
\begin{align}
E_{\mathrm H, \mathrm S}[P] &= \frac{1}{2} \sum_{\mathbf k} \Tr \left\lbrace P h_{\mathrm H}^{\mathrm S}[P - \mathbb{1}_4] \right\rbrace \\
&= \frac{1}{2A} \sum_{\mathbf G, \mathbf k, \mathbf k^\prime} V_{\mathbf G} \Tr \left\lbrace P \Lambda_{\mathrm S}(\mathbf k, \mathbf G) \right\rbrace 
\nonumber \\ &\quad \times
\Tr\left\lbrace (P - \mathbb{1}_4) \Lambda_{\mathrm S}^\dagger(\mathbf k^\prime, \mathbf G) \right\rbrace.
\end{align}
For the magnetically ordered phases, substituting the density matrices gives
\begin{align}
E_{\mathrm H, \mathrm S}[P_{\mathrm{FM}, \mathrm{AFM}}] = -\Delta E_{\mathrm S}
\end{align}
and hence $E_{\mathrm{HF}}[P_{\mathrm{FM}, \mathrm{AFM}}] = 0$, meaning these density matrices describe ground states of $\mathcal{H}_S$.
The Hartree contribution for the charge-density-wave state evaluates to
\begin{align}
E_{\mathrm{H}}[P_{\mathrm{CDW}}] = - \frac{2}{A} \sum_{\mathbf G} V_{\mathbf G} \left(\sum_{\mathbf k}   F_{\mathrm S}(\mathbf k, \mathbf G) \rme^{\rmi \Phi_{\mathrm S}(\mathbf k, \mathbf G)}\right)^2,
\end{align}
yielding a total energy
\begin{align}
E_{\mathrm{HF}}[P_{\mathrm{CDW}}] = \frac{2}{A} \sum_{\mathbf G} V_{\mathbf G} \left(\sum_{\mathbf k}   F_{\mathrm S}(\mathbf k, \mathbf G) \sin \Phi_{\mathrm S}(\mathbf k, \mathbf G)\right)^2,
\end{align}
where we have used Eqs.~\eqref{eq:trs_f}, \eqref{eq:trs_phi}, and~\eqref{eq:lambda-dagger-periodic} to simplify the result.
Since neither of the symmetry restrictions in Eqs.~\eqref{eq:trs_f} and~\eqref{eq:trs_phi} force the inner sum to vanish, the charge-density-wave state will generically not be a ground state of $\mathcal H_{\mathrm S}$.

The calculation of the antichiral contribution $\mathcal H_{\mathrm A}$ proceeds analogously, leading to the results presented in the main text.

\subsection{Effect of dispersion}
\label{app:effect-of-dispersion}

In the following, we perturbatively analyze the effect of the renormalized kinetic energy term
\begin{align}
\mathcal H_0 = \sum_{\mathbf k} c^\dagger_{\mathbf k} \tilde h(\mathbf k) c_{\mathbf k}
\end{align}
in the two-band model [Eq.~\eqref{eq:strong-coupling-model}] on the energy competition between the magnetic orderings.
The first-order energy correction turns out to be identical for both phases, but the second-order contribution favors the antiferromagnetic ground state.
Intuitively, this can be understood as an analogue of the antiferromagnetic Heisenberg exchange interaction with coupling $J \propto t^2/U$ that arises in the strong-coupling expansion of the Hubbard model.

Note that the $C_{2z}\mathcal  T$ and SU(2) symmetries imply the form
\begin{align}
\tilde h(\mathbf k) = ( a(\mathbf k) \sigma_0 + b(\mathbf k) \sigma_x \rme^{\rmi \Phi_0(\mathbf k) \sigma_z}) s_0
\end{align}
for the matrix $\tilde h(\mathbf k)$, where $a$, $b$ and $\Phi_0$ are real-valued functions of the wavevector $\mathbf k$.
The four-dimensional ground-state subspace $\mathcal D$ of the chiral Hamiltonian $\mathcal H_\mathrm{S}$ is spanned by the states
\begin{align}
\ket{\text{AFM}_1} &= \prod_{\mathbf k} c^\dagger_{\mathbf k \mathrm{A} \uparrow} c^\dagger_{\mathbf k \mathrm{B} \downarrow} \ket 0, \\
\ket{\text{AFM}_2} &= \prod_{\mathbf k} c^\dagger_{\mathbf k \mathrm{A} \downarrow} c^\dagger_{\mathbf k \mathrm{B} \uparrow} \ket 0, \\
\ket{\text{FM}_1} &= \prod_{\mathbf k} c^\dagger_{\mathbf k \mathrm{A} \uparrow} c^\dagger_{\mathbf k \mathrm{B} \uparrow} \ket 0, \\
\ket{\text{FM}_2} &= \prod_{\mathbf k} c^\dagger_{\mathbf k \mathrm{A} \downarrow} c^\dagger_{\mathbf k \mathrm{B} \downarrow} \ket 0,
\end{align}
which diagonalize the perturbation $\mathcal H_0$ on $\mathcal D$ and thus represent a suitable basis for the degenerate perturbation theory.
The corresponding energy corrections to first order in $\mathcal H_0$ are given by
\begin{align}
\bra*{\text{AFM}_{i}} \mathcal H_0 \ket*{\text{AFM}_{i}} &= \bra*{\text{FM}_{i}} \mathcal H_0 \ket*{\text{FM}_{i}} 
= 2 \sum_{\mathbf k} a(\mathbf k),
\end{align}
for $i=1,2$,
and thus do not distinguish between the two different magnetic orderings.
We therefore consider the second-order contribution to the energy of each of the four basis states $\ket \Psi$, given by \cite{sakuraibook}
\begin{align}
\Delta^{(2)}(\ket \Psi) &= \bra{\Psi} \mathcal{H}_0 \frac{\phi_\mathcal D}{\bra{\Psi} \mathcal H_\mathrm{S} \ket{\Psi} - \mathcal H_{\mathrm{S}}} \mathcal H_0 \ket{\Psi} \\
&= -\sum_{\ket \Phi, \ket \Theta \not \in \mathcal D} \bra \Psi \mathcal H_0 \ket \Phi \bra \Phi \mathcal H_{\mathrm S}^{-1} \ket \Theta \bra \Theta \mathcal H_0 \ket \Psi .
\end{align}
Here, $\phi_\mathcal{D} = \mathbb{1} - \sum_{i=1, 2} (\ket{\text{AFM}_i}\bra{\text{AFM}_i} + \ket{\text{FM}_i}\bra{\text{FM}_i})$ projects into the complement of the ground-state subspace, and $\ket \Phi$, $\ket \Theta$ are Slater determinants forming a basis of this complement.
Note that $\mathcal H_0$ is diagonal in momentum and spin, and hence the only Slater determinant states that can contribute finite matrix elements $\bra \Psi \mathcal H_0 \ket \Phi$ to the above sum are of the form $\ket \Phi = c^\dagger_{\mathbf k \sigma s} c_{\mathbf k \sigma^\prime s} \ket \Psi$, where $c^\dagger_{\mathbf k \sigma s}$ creates an electron with spin $s$ in sublattice $\sigma$, and $\sigma \neq \sigma^\prime$ to ensure that $\ket \Phi \not \in \mathcal D$.
However, applying the operator $c^\dagger_{\mathbf k \sigma s} c_{\mathbf k \sigma^\prime s}$ (with $\sigma \neq \sigma^\prime$) to a ferromagnetic state will necessary annihilate it, since both sublattices are already occupied with the same spin species.
Hence, the sum collapses and the energy correction evaluates to 
\begin{align} \label{eq:Delta2-FM}
\Delta^{(2)}(\ket{\text{FM}_i}) = 0, \quad i=1,2.
\end{align}

For the antiferromagnetic states, we instead find
\begin{align} \label{eq:Delta2-AFM}
\Delta^{(2)}(\ket*{\text{AFM}_1}) &= - \sum_{\mathbf k, \mathbf k^\prime, m, m^\prime} \bra*{\text{AFM}_1} \mathcal H_0 \ket*{\Phi^{\mathrm{eh}}_{\mathbf k, m}}
\nonumber \\ & \times
\bra*{\Phi^{\mathrm{eh}}_{\mathbf k, m}} \mathcal H_{\mathrm{S}}^{-1} \ket*{\Phi^{\mathrm{eh}}_{\mathbf k^\prime, m^\prime}} \bra*{\Phi^{\mathrm{eh}}_{\mathbf k^\prime, m^\prime}} \mathcal H_0 \ket*{\text{AFM}_1},
\end{align}
with electron-hole pair states given by
\begin{align}
\ket*{\Phi^{\mathrm{eh}}_{\mathbf k, m}} =
\begin{cases}
c^\dagger_{\mathbf k \mathrm{A} \downarrow} c_{\mathbf k \mathrm{B} \downarrow} \ket{\text{AFM}_1} & \text{for } m = \mathrm{A}, \\
c^\dagger_{\mathbf k \mathrm{B} \uparrow} c_{\mathbf k \mathrm{A} \uparrow} \ket{\text{AFM}_1}  & \text{for } m = \mathrm{B},
\end{cases},
\end{align}
and
\begin{align}
\bra*{\text{AFM}_1} \mathcal H_0 \ket*{\Phi^{\mathrm{eh}}_{\mathbf k, m}} = b(\mathbf k)\rme^{-\rmi \Phi_0(\mathbf k)} \delta_{m \mathrm{A}} + b(\mathbf k)\rme^{\rmi \Phi_0(\mathbf k)} \delta_{m \mathrm{B}}.
\end{align}
Defining
$v_{\mathbf k, m} = \bra*{\Phi^{\mathrm{eh}}_{\mathbf k, m}} \mathcal H_0 \ket*{\text{AFM}_1}$
and 
$M_{\mathbf k, m; \mathbf k^\prime m^\prime} = \bra*{\Phi^{\mathrm{eh}}_{\mathbf k, m}} \mathcal H_{\mathrm{S}} \ket*{\Phi^{\mathrm{eh}}_{\mathbf k^\prime, m^\prime}}$,
Eq.~\eqref{eq:Delta2-AFM} can be written in the form
\begin{align}
\Delta^{(2)} (\ket{\text{AFM}_1}) = 
-\sum_{\mathbf k, \mathbf k', m, m'} v^\dagger_{\mathbf k,m} M^{-1}_{\mathbf k,m; \mathbf k',m'} v_{\mathbf k',m'}.
\end{align}
An analogous relation holds for $\Delta^{(2)}(\ket{\mathrm{AFM}_2})$ and, by symmetry, yields the same energy correction.
The chiral Hamiltonian $\mathcal H_{\mathrm S}$ is positive semidefinite. Assuming the ferromagnetic and antiferromagnetic states exhaust the ground-state subspace of $\mathcal H_{\mathrm S}$, it follows that the matrix $M$ has strictly positive eigenvalues. Consequently, the second-order correction to the energy of the antiferromagnetic state is strictly negative, $\Delta^{(2)} (\ket{\text{AFM}_i})<0$ for $i=1,2$.
Together with the fact that the second-order contribution to the ferromagnetic state vanishes [Eq.~\eqref{eq:Delta2-FM}], this demonstrates that the kinetic term favors antiferromagnetic ordering, as stated in the main text.

\appsection{Effective permittivity}
\label{app:permittivity}

This appendix provides a theoretical estimate for the effective permittivity $\epsilon_{\mathrm{eff}}(2.7^\circ)$ at the critical twist angle, showing that values $\epsilon_\text{eff} \sim \mathcal O(100)$ are indeed realistic for twisted double bilayer WSe$_2$.
We first evaluate the momentum-dependent dielectric function $\epsilon(\mathbf q)$ within the random phase approximation, and subsequently calculate an effective constant permittivity $\epsilon_\text{eff}$ in terms of a suitably weighted average of $\epsilon(\mathbf q)$.

Following Ref.~\cite{goodwin19}, we assume that the diagonal elements of the dielectric tensor dominate and can be evaluated within the random phase approximation as
\begin{align}
\epsilon(\mathbf q) = \epsilon_{\mathrm{env}} + v(\mathbf q) \Pi_0(\mathbf q).
\end{align}
Here, $\epsilon_{\mathrm{env}}$ accounts for the substrate contribution, $v(\mathbf q) = e^2 / 2\epsilon_0 \vert \mathbf q\vert$ is the bare Coulomb potential, and the independent-particle polarizability $\Pi_0$ is given by~\cite{goodwin19}
\begin{align}
\Pi_0(\mathbf q) = \frac{4}{A} \sum_{\mathbf k \in \mathrm{mBZ}} \sum_{\mathrm c, \mathrm v} \frac{\big\vert \bra{\psi_{\mathrm v, \mathbf k}} \rme^{-\rmi\mathbf q \cdot \mathbf r} \ket{\psi_{\mathrm c, \mathbf k + \mathbf q}} \big\vert^2}{\varepsilon_{\mathrm c, \mathbf k + \mathbf q} - \varepsilon_{\mathrm v, \mathbf k}},
\end{align}
where $\varepsilon_{\mathrm c,\mathbf k}$ and $\varepsilon_{\mathrm v,\mathbf k}$ denote the conduction and valence band dispersions at hole doping of $\nu = 2$ and $\ket{\psi_{\mathrm c,\mathbf k}}$ and $\ket{\psi_{\mathrm v,\mathbf k}}$ label the corresponding Bloch states. 
The matrix elements of the operator $\rme^{-\rmi\mathbf q\cdot \mathbf r}$ are evaluated using the expression in Eq.~\ref{eq:bloch-states} for the Bloch states, giving
\begin{align}
\bra{\psi_{\mathrm v, \mathbf k}} \rme^{-\rmi\mathbf q \cdot \mathbf r} \ket{\psi_{\mathrm c, \mathbf k + \mathbf q}} = \sum_{\mathbf G, \ell} u^*_{\mathbf k, \mathrm{v};\mathbf G, \ell} u_{\mathbf k + \mathbf q, \mathrm{c};\mathbf G, \ell}.
\end{align}

To facilitate the theoretical modeling, we further decompose the dielectric function as
\begin{align} \label{eq:epsilon-q}
\epsilon(\mathbf q) = \epsilon_{\mathrm{env}} + v(\mathbf q) (\Pi_0^{\mathrm{cRPA}}(\mathbf q) + \Pi_0^{\mathrm{flat}}(\mathbf q)),
\end{align}
where $\Pi_0^{\mathrm{cRPA}}(\mathbf q)$ accounts for transitions involving remote bands, and
\begin{align} \label{eq:pi0-flat}
\Pi_0^{\mathrm{flat}}(\mathbf q) = \frac{4}{A} \sum_{\mathbf k \in \mathrm{mBZ}} \frac{\big\vert \bra{\psi_{-, \mathbf k}} \rme^{-\rmi\mathbf q \cdot \mathbf r} \ket{\psi_{+, \mathbf k + \mathbf q}} \big\vert^2}{\varepsilon_{+, \mathbf k + \mathbf q} - \varepsilon_{-, \mathbf k}},
\end{align}
arises from transitions within the two nearly-flat graphene-like bands $\varepsilon_{\pm, \mathbf q}$.
From previous calculations in the context of twisted bilayer graphene~\cite{pizarro19, goodwin19, biedermann25}, we expect the flat-band contribution $\Pi_0^\text{flat}$ to yield the dominant contribution to the effective permittivity. It can be directly computed within the low-energy continuum approximation.
Accurately computing the remote-band contribution $\Pi_0^\text{cRPA}$ requires to go beyond the continuum approximation, as also remote bands that are not well captured within the continuum model may contribute to the effective permittivity.
Here, we estimate this contribution from the permittivity of an uncoupled WSe$_2$ tetralayer, $v(\mathbf q) \Pi_0^{\text{cRPA}}(\mathbf q) \simeq g_{\mathrm{l}} (\epsilon_{\text{mono-WSe$_2$}} - 1)$, where $\epsilon_\text{mono-WSe$_2$} \simeq 2.9$ is the in-plane static permittivity of a WSe$_2$ monolayer~\cite{kumar12}, $g_{\mathrm{l}} = 4$ counts the number of layers, and we subtract one to prevent double counting of the vacuum permittivity.
This assumption neglects the increase of screening that arises from the flattening of the remote bands for small twist angles and may thus be understood as a lower bound for the corresponding contribution from $\Pi_0^\text{cRPA}$ to $\epsilon(\mathbf q)$~\cite{pizarro19, goodwin19}.
In passing, we note that a meaningful comparison of the critical interaction strength in twisted double bilayer TMDs with the critical value $(U_0/t_1)_\mathrm{c} \simeq 3.8$ for the honeycomb-lattice Hubbard model~\cite{assaad13} requires accounting for the screening captured by $\Pi_0^\text{cRPA}$. Including this effect substantially lowers the critical Wigner-Seitz radius relative to the estimate given in Ref.~\cite{ma24}.

\begin{figure}[tb!]
\includegraphics[width=\linewidth]{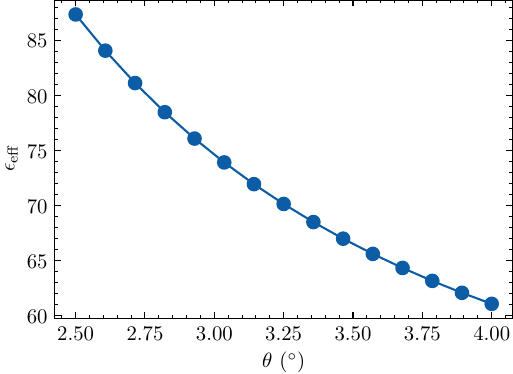}
\caption{%
Effective permittivity $\epsilon_{\mathrm{eff}}$ as a function of the twist angle $\theta$ in the Dirac semimetal phase for a substrate permittivity $\epsilon_{\mathrm{env}} = 5$. Here, a $17 \times 17$ momentum-space grid is used to avoid singularities in the momentum sum of Eq.~\eqref{eq:pi0-flat}.
At the experimentally determined critical twist angle $\theta_{\mathrm c} = 2.7^\circ$, we find $\epsilon_{\mathrm{eff}}(\theta_{\mathrm c}) \simeq 82$.}
\label{fig:10}
\end{figure}

An effective dielectric constant suitable as an input for the Hartree-Fock analysis can be obtained by averaging the momentum-dependent dielectric function $\epsilon(\mathbf q)$ such that the interaction energy scale $U$ is preserved. In analogy with Appendix~\ref{app:energy-scales}, we define
\begin{align}
U =\frac{1}{2AL^2}  \sum_{\mathbf q, \mathbf k} \tilde V_{\mathbf q} F(\mathbf k, \mathbf q)^2\,,
\end{align}
with $\tilde V_{\mathbf q} = \epsilon_{\mathrm{eff}} V_{\mathbf q} / \epsilon(\mathbf q)$ the Coulomb potential with momentum-dependent permittivity, and $F(\mathbf k, \mathbf q)^2 = \Tr \left[ \Lambda(\mathbf k, \mathbf q) \Lambda^\dagger(\mathbf k, \mathbf q) \right] / 4$ the magnitude of the overlap matrices $\Lambda(\mathbf k, \mathbf q)$ in the two-band model.
The energy scale $U$ is preserved when replacing the Coulomb interaction with momentum-dependent permittivity by an effective Coulomb potential with dielectric constant $\epsilon_\text{eff}$, i.e., $\tilde V_\mathbf{q}  \mapsto V_\mathbf q$, provided that
\begin{align}
\epsilon_{\mathrm{eff}} = \frac{\sum_{\mathbf q, \mathbf k} \epsilon(\mathbf q) \tilde V_{\mathbf q} \, F(\mathbf k, \mathbf q)^2}{\sum_{\mathbf q, \mathbf k} \tilde V_{\mathbf q} \, F(\mathbf k, \mathbf q)^2}\,.
\end{align}
Numerical results for the resulting effective dielectric constant are shown in Fig.~\ref{fig:10}.
At the experimentally determined critical twist angle $\theta_{\mathrm c} = 2.7^\circ$, we find $\epsilon_{\mathrm{eff}}(\theta_{\mathrm c}) \simeq 82$, close to the semimetal-to-insulator phase boundary shown in Fig.~\ref{fig:4}(c).
We note that the above calculation is based on the noninteracting continuum Hamiltonian assuming a Dirac semimetal ground state. Interactions may modify the effective permittivity. This effect is expected to be particularly significant in the presence of a finite interaction-induced band gap, which is why we only present results for twist angles above approximately $\theta_\mathrm{c}$.

\appsection{Critical pressure for larger twist angle}
\label{app:pressure}

\begin{figure}[h!]
\includegraphics[width=\linewidth]{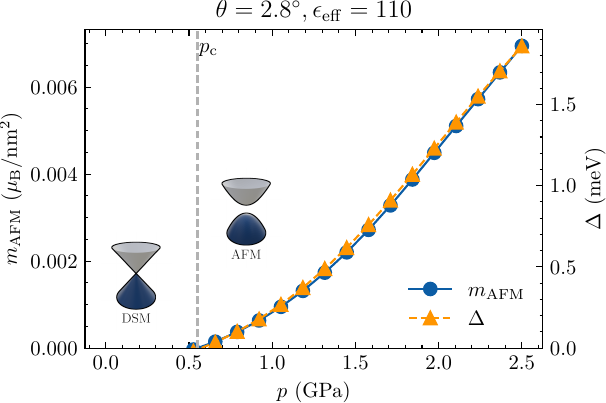}
\caption{%
Same as Fig.~\ref{fig:5}, but for slightly larger twist angle $\theta = 2.8^\circ$.
The continuous transition from the symmetric Dirac semimetal to the antiferromagnetic insulator occurs at a critical pressure $p_\mathrm{c} \simeq 0.6\,\mathrm{GPa}$.
}
\label{fig:11}
\end{figure}

This appendix presents additional data on the pressure-induced transition for a sample with a slightly larger twist angle than in Fig.~\ref{fig:5}.
Figure~\ref{fig:11} shows the staggered magnetization density $m_\text{AFM}$ and the spectral gap $\Delta$, extrapolated to the thermodynamic limit, as functions of pressure $p$ for $\theta = 2.8^\circ$. The results highlight the sensitivity of the critical pressure to the twist angle: For $\theta - \theta_\mathrm{c}(0) = 0.05^\circ$, the semimetal-to-insulator transition occurs at $p_\mathrm{c} \simeq 0.2\,\text{GPa}$ (cf.~Fig.~\ref{fig:5}), whereas for $\theta - \theta_\mathrm{c}(0) = 0.10^\circ$, it shifts to $p_\mathrm{c} \simeq 0.6\,\text{GPa}$.

\FloatBarrier
\bibliographystyle{longapsrev4-2}
\bibliography{tdbWSe2}

\end{document}